\documentclass[iop]{emulateapj}
\usepackage{amsmath}
\usepackage{color}
\usepackage{ulem}
\usepackage{tablefootnote}

\def\ltsima{$\; \buildrel < \over \sim \;$}
\def\simlt{\lower.5ex\hbox{\ltsima}}
\def\gtsima{$\; \buildrel > \over \sim \;$}
\def\simgt{\lower.5ex\hbox{\gtsima}}

\slugcomment{Draft version -----, 2017}

\shorttitle{Broad-band X-ray analysis of Mrk 463}
\shortauthors{Yamada et al.}

\begin{document}

\title{Broadband X-ray Spectral Analysis of the Double-Nucleus Luminous
Infrared Galaxy Mrk 463}

\author{
Satoshi   Yamada\altaffilmark{1},
Yoshihiro Ueda\altaffilmark{1},
Saeko     Oda\altaffilmark{1},
Atsushi   Tanimoto\altaffilmark{1},
Masatoshi Imanishi\altaffilmark{2,3}, \\\
Yuichi    Terashima\altaffilmark{4}, and
Claudio   Ricci\altaffilmark{5,6,7}
}
\affil
{
\altaffilmark{1}Department of Astronomy, Kyoto University, Kitashirakawa-Oiwake-cho, Sakyo-ku, Kyoto 606-8502, Japan; styamada@kusastro.kyoto-u.ac.jp \\
\altaffilmark{2}National Astronomical Observatory of Japan, 2-21-1 Osawa, Mitaka, Tokyo 181-8588, Japan \\
\altaffilmark{3}Department of Astronomy, School of Science, SOKENDAI (The Graduate University for Advanced Studies), 2-21-1 Osawa, Mitaka, Tokyo 181-8588, Japan \\
\altaffilmark{4}Department of Physics, Ehime University, Matsuyama 790-8577, Japan\\
\altaffilmark{5}N\'ucleo de Astronom\'ia de la Facultad de Ingenier\'ia, Universidad Diego Portales, Av. Ej\'ercito Libertador 441, Santiago, Chile\\
\altaffilmark{6}Kavli Institute for Astronomy and Astrophysics, Peking University, Beijing 100871, China\\
\altaffilmark{7}Chinese Academy of Sciences South America Center for Astronomy, Camino El Observatorio 1515, Las Condes, Santiago, Chile
}

\begin{abstract}

We present a broadband (0.4--70 keV) X-ray spectral analysis of the
luminous infrared galaxy (LIRG) system Mrk 463 observed with 
\textit{Nuclear Spectroscopic Telescope Array} (\textit{NuSTAR}),
\textit{Chandra}, and \textit{XMM-Newton}, which contains double active
galactic nuclei (AGNs) (Mrk 463E and Mrk 463W) with a separation of
$\sim$ 3.8 kpc. Detecting their transmitted hard X-ray $>10$ keV
continua with \textit{NuSTAR},  
we confirm that Mrk~463E and Mrk~463W have AGNs with
intrinsic luminosities of (1.6--2.2) $\times 10^{43}$ and (0.5--0.6)
$\times 10^{43}$ erg s$^{-1}$ (2--10 keV) obscured by hydrogen column
densities of 8 $\times 10^{23}$ and 3 $\times 10^{23}$ cm$^{-2}$,
respectively. Both nuclei show strong reflection components from cold
matter. The luminosity ratio between X-ray (2--10 keV) and [O IV] 25.89
$\mu$m of Mrk 463E is $\sim$5 times
smaller than those of normal Seyfert galaxies, 
suggesting that the intrinsic SED is
X-ray weak relative to the UV
luminosity. In fact, the bolometric AGN luminosity of Mrk 463E estimated
from $L^{\prime}$-band (3.8 $\mu$m), [O IV] 25.89~$\mu$m, and
[Ne~V] 14.32~$\mu$m lines indicate a large bolometric-to-X-ray
luminosity ratio, $\kappa_{\rm 2-10 keV} \approx$ 
110--410, and a high Eddington ratio, $\lambda_{\rm
Edd} \sim$ 0.4--0.8. We suggest that the merger triggered a rapid
growth of the black hole in Mrk 463E, which is not yet deeply ``buried''
by circumnuclear dust.
By contrast, the $L^{\prime}$-band luminosity of Mrk
463W is unusually small relative to the X-ray luminosity, suggesting
that the Eddington ratio is low ($<10^{-3}$) and it
might be still in an early phase of merger-driven AGN activity.

\end{abstract}


\section{Introduction}

It is a widespread belief that supermassive black holes (SMBHs) and
galactic bulges have co-evolved by regulating each other's growth
\citep[see e.g.,][for a recent review]{Kormendy2013}.
A key mechanism driving the co-evolution is mergers of gas-rich galaxies,
which trigger 
intense star formation (SF) and rapid mass accretion onto the
SMBHs,which, during this phase, are observed as active galactic nuclei (AGNs)
\citep[e.g.,][]{Mihos1996,Struck1999,Hopkins2008}.
Dust-rich merging galaxies emit the bulk of their bolometric luminosities
in the infrared, and are observed as luminous ($L_{\rm IR}$(8--1000 $\mu$m) = $10^{11}$--$10^{12}
L_{\odot}$) or ultra-luminous ($L_{\rm IR} \geq 10^{12}L_{\odot}$)
infrared galaxies (LIRGs and ULIRGs; \citealt{Sanders1996}).
As the central regions of these objects are deeply obscured by gas and
dust, multiwavelength observations are required to understand the 
nature of the system, 
including the main energy source (SF or AGN) and the structure of the nucleus.
In particular, thanks to their strong penetrating power against
obscuration, hard X-ray ($>10$ keV) observations are very useful to
detect hidden AGNs, even Compton-thick ones with column densities 
of log~$N_{\rm H}$/cm$^{-2}$ = 24--25. 
Following earlier studies at energies below 10 keV with 
\textit{XMM-Newton} \citep[e.g.,][]{Imanishi2003,Pereira2011} and \textit{Chandra} \citep[e.g.,][]{Teng2005,Iwasawa2011}, 
the \textit{Nuclear Spectroscopic Telescope Array} (\textit{NuSTAR}), the first
hard X-ray imaging mission covering the 3--79~keV range 
\citep{Harrison2013}, is now starting
to unveil the nature of heavily obscured AGNs in local U/LIRGs by 
directly detecting their transmitted component
\citep[e.g.,][]{Teng2015,Ricci2016,Oda2017,Nardini2017}.

To understand the role of mergers in feeding SMBHs, it is important to
investigate the AGN properties as a function of the separation between
the two galaxies. In this context, studies of dual AGN systems
\citep[e.g.,][]{Komossa2003,Koss2011,Fabbiano2011} are of particular
interest, since they are considered to be in an earlier stage of mergers than
typical U/LIRGs showing highly compact cores. From the 22 and 58 month \textit{Swift}/BAT
hard X-ray survey catalog, \citet{Koss2012} identified 17 dual AGNs. They found 
that the AGN luminosities increases with decreasing
separation. Using \textit{NuSTAR} observations of local merging galaxies, \citet{Ricci2017}
showed that the degree of AGN obscuration increases with the evolutionary
stage of mergers; the majority of AGNs in the late-merger galaxies are
subject to heavy obscuration for which galactic-scale gas and dust
could be responsible. 
Detailed studies of AGN structure in {\it individual} dual AGNs based on multiwavelength 
data could help us to better understand the physical processes induced by mergers.
However, currently the number of sources for which this has been possible is very limited.

Mrk 463 ($z = 0.0504$) is a merging system consisting of Mrk 463E (east)
and Mrk 463W (west), which show tidal tails as a result of the
galaxy-galaxy interaction \citep{Mazzarella1991}. This is a dual AGN
system with the third closest physical separation (3.8 kpc, corresponding to
an angular separation of 3.8'') following NGC 6240 (1.5 kpc) and Mrk 739 (3.4 kpc)
among the \textit{Swift}/BAT dual-AGN sample compiled by \citet{Koss2012}, and hence is
an ideal target to link earlier and later stage mergers. Up to present,
detailed X-ray spectral analyses have been done only 
below 10 keV. \citet{Imanishi2004} observed it with \textit{XMM-Newton} and found that the
X-ray spectrum of Mrk 463 is heavily obscured with $N_{\rm H} 
\sim 3 \times 10^{23}$ cm$^{-2}$ showing an iron-K$\alpha$ emission line at 6.4 keV.
However, the
two nuclei were not separated due to the limited angular
resolution. Using \textit{Chandra}, \citet{Bianchi2008} detected AGNs from both Mrk 463E
and 463W, confirming that both spectra are heavily obscured. The Mrk 463
system is classified as a LIRG on the basis of the total infrared
luminosity, $L_{\rm IR}$(8--1000 $\mu$m) = 10$^{11.8}L_{\odot}$ \citep{Imanishi2014}. In
the mid-IR band, Mrk 463E is notably brighter than Mrk~463W \citep{Armus2004}. The
\textit{Spitzer}/IRS spectrum of Mrk 463E was reported in \citet{Armus2004}, where high
excitation lines such as [O IV] 25.89 $\mu$m and [Ne V] 14.32 $\mu$m were detected.

In this paper, we perform a broadband X-ray (0.4--70~keV) spectral
analysis of Mrk 463, by simultaneously utilizing the data from \textit{NuSTAR},
\textit{XMM-Newton}, and \textit{Chandra}. 
Detecting their transmitted hard X-ray ($>10$ keV)
continua with \textit{NuSTAR}, which are little affected by obscuration,
and separating the soft X-ray ($<8$ keV) spectra of Mrk 463E and Mrk
463W with \textit{Chandra},
we determine the
intrinsic luminosities and absorption column densities of both galaxies
with the best accuracy so far, 
by taking into account 
the reflection components from circumnuclear matter.
This paper is structured as follows. Section~\ref{sec2} describes the
X-ray observations and data reduction. In Section~\ref{sec3}, we present
our spectral analysis using two models: an analytical model and a
Monte-Carlo based torus model by \citet{Ikeda2009}. In
Section~\ref{sec4}, we discuss our results mainly by comparing our
results with the mid-IR properties. Section~\ref{sec5} summarizes the
conclusions. We adopt the cosmological parameters of ($H_0$ ,
$\Omega_{\rm m}$ , $\Omega_{\rm \Lambda}$) = (70 km s$^{-1}$ Mpc$^{-1}$
, 0.3, 0.7) and the solar abundances of \citet{Anders1989} throughout
this paper. All errors of the spectral parameters are quoted at 90\%
confidence limits for a single parameter of interest.

\section{Observations and Data Reduction}
\label{sec2}

The observation log of the X-ray data used in this paper (\textit{NuSTAR}, \textit{Chandra},
\textit{XMM-Newton}) is given in Table~\ref{tab1-obs}. Details of the data reduction are
described below.

\begin{deluxetable*}{ccrrc}
\tablewidth{\textwidth}

\tablecaption{Observation Log of Mrk 463 Utilized in this Work \label{tab1-obs}}
\tablehead{
\colhead{Satellite}      &
\colhead{ObsID}          &
\colhead{Start Date(UT)} &
\colhead{End Date(UT)}   &
\colhead{Net exp.(ks)} 
}
\startdata
\textit{NuSTAR} & 60061249002 & 2014-01-01 22:31:07 & 2014-01-02 10:41:07 & 23.9 \\
\textit{Chandra} (ASIS-S) & 4913 & 2004-06-11 23:06:47 & 2004-06-12 13:42:31 & 48.8 \\
\textit{XMM-Newton} & 0094401201 & 2001-12-22 03:13:30 & 2001-12-22 10:40:18 & 25.8 
\enddata

\end{deluxetable*}

\subsection{NuSTAR}

\textit{NuSTAR} observed Mrk 463 on 2014 January 1 for a net exposure of 23.9 ks. 
\textit{NuSTAR} carries two co-aligned grazing
incidence telescopes coupled with two focal plane modules A and B (FPMA
and FPMB), which are sensitive to X-ray energies of 3--79 keV. 
The angular resolution of the telescopes is 58'' in a half-power
diameter (HPD). The data were reduced by using the standard pipeline script
(\textsc{nupipeline}) available in the \textit{NuSTAR} Data Analysis Software (NuSTARDAS,
v1.6.0; part of the HEASOFT distribution as of v6.19) with \textit{NuSTAR}
Calibration Database (CALDB) v20161021. Since Mrk~463E and Mrk~463W cannot be separated by the
point spread function of \textit{NuSTAR}, we produced the total
spectrum of the Mrk 463 system by accumulating photon events in a
circular region with a radius of 50'' centered around the two
galaxies.
The background was taken from a source-free circular region with 
a radius of 60''.  The spectra
and light curves were extracted with the \textsc{nuproducts} task. 
Since the signal-to-noise ratio of the FPMA and FPMB spectra are limited, 
these spectra were then combined by using \textsc{addascaspec} 
to improve the statistics. 
The spectra were binned to contain a minimum of 50
counts per bin.
We tested against the null hypothesis that the 3--24 keV flux is constant via a $\chi^2$ test.
No significant time variability on a timescale of hours was found at $>90\%$ confidence limits.

\subsection{Chandra}

\begin{figure}
    \centering
    \includegraphics[keepaspectratio,scale=0.45]
    {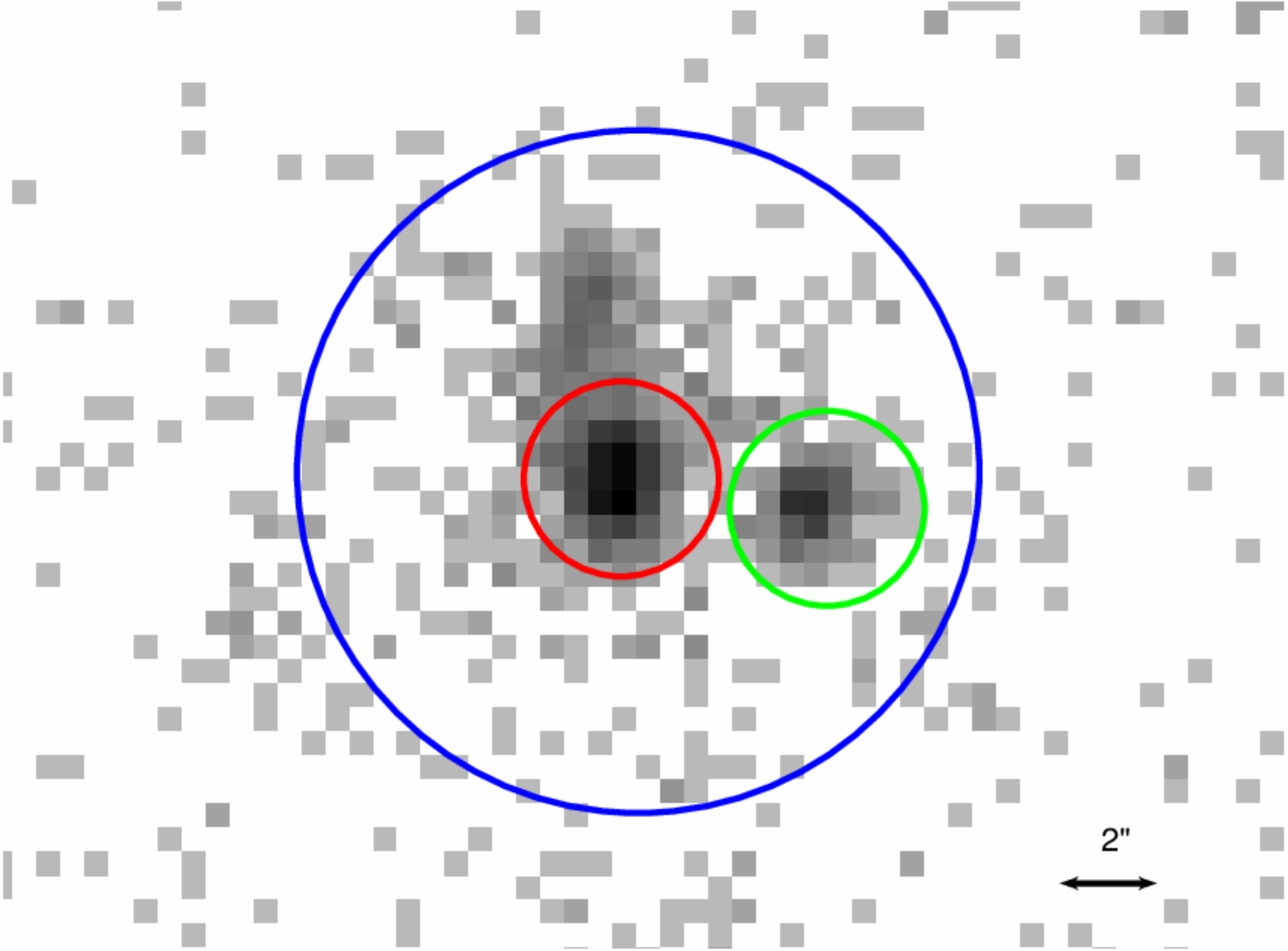}
    \caption{\textit{Chandra}/ASIS-S image (0.4-8 keV) of the central region of
 Mrk 463. The 3 spectral extraction regions are
 shown by the circles for Mrk 463E (red), Mrk 463W (green), and the
 extended emission (blue).
    }
    \label{fig1-Mrk463image}
\end{figure}

\textit{Chandra} \citep{Weisskopf2002}
observed Mrk 463 on 2004 June 11 with the Advanced CCD Imaging Spectrometer
(ACIS: \citealt{Garmire2003}) for a net exposure of 50.0 ks. 
Thanks to its excellent angular resolution ($<1$''), \textit{Chandra} 
was able to spatially resolve 
two galaxies as well as extended X-ray emission surrounding them \citep{Bianchi2008}. 
We reduced the data using
\textit{Chandra} Interactive Analysis of Observations (CIAO)
v4.8.1 and the Calibration Database (CALDB) v4.7.2. 
The event files were reprocessed with the CIAO \textsc{chandra\_repro} script. 
Following \citet{Bianchi2008}, we 
used three extraction regions: 
two circular regions with a radius of 2'' centered
on the nuclei of Mrk~463E and Mrk~463W, 
and one circular region with a radius of 7'' 
encompassing all soft X-ray extended emission from which 
the two nucleus regions are excluded.
The image in the 0.4-8 keV band is shown in Figure~\ref{fig1-Mrk463image}.
The spectra, the redistribution matrix function (RMF) and the auxiliary response file (ARF) 
were produced with the task \textsc{specextract}. 
The spectral bins were grouped to contain no less than 25,
20, and 20 counts for Mrk 463E, Mrk 463W, and the soft extended
emission, respectively, so that $\chi^2$ statistics is applicable while
line features were not smeared out beyond the energy resolution.

\subsection{XMM-Newton}

An \textit{XMM-Newton} \citep{Jansen2001}
observation of Mrk 463 was performed on 2001 December 22 for a net exposure of 26.8 ks.
The main instruments of \textit{XMM-Newton} are three European Photon Imaging
Cameras (EPICs): PN \citep{Struder2001}, MOS1, and MOS2 \citep{Turner2001}.
The data were reduced in a standard manner 
with the \textit{XMM-Newton} Science Analysis System (SAS) v15.0.0. We removed 
events recorded during background flares, and only utilized
those with patterns 0-4 for PN and patterns 0-12 for MOS. 
The spectra of Mrk 463E, Mrk 463W, and the soft extended emission 
cannot be separated with \textit{XMM-Newton}, whose angular resolution is 
15'' in HPD. The
total source spectra was extracted from a circular region with a radius of
28'' centered at the middle point between the two galaxies, while the
background was taken from a nearby circular region 
with a radius of 50''. The RMF and ARF of each EPIC
were generated by using the tasks \textsc{rmfgen} and \textsc{arfgen}. The MOS1 and MOS2 spectra
were combined into a single spectrum with \textsc{addascaspec}.  
The MOS1+MOS2 and PN spectra were binned to contain more than 30~counts per
bin.

\section{X-ray Spectral Analysis}
\label{sec3}

To best constrain the broadband spectral properties of Mrk 463, we
simultaneously analyze six spectra: those of Mrk 463E, Mrk
463W, and the extended emission obtained with \textit{Chandra}/ACIS-S (0.4--7~keV, 0.6--8~keV, 0.4--1.6~keV),
and the total spectra (including all emission from Mrk 463) obtained
with \textit{NuSTAR}/FPMs (3--70~keV), \textit{XMM-Newton}/EPIC-MOSs (0.47--11~keV), and
\textit{XMM-Newton}/EPIC-PN (0.49--12~keV), where the highest signal-to-noise ratios
are achieved. 
The spectra folded with the energy responses are plotted in 
Figure~\ref{fig2-fitting}.
We consider models independently for the three regions
(Mrk~463E, Mrk~463W, and Extended) and apply each model to the
corresponding \textit{Chandra} spectrum, while the \textit{NuSTAR} and \textit{XMM-Newton} spectra
are simultaneously fit with their summed model.  Spectral fitting is
performed on XSPEC \citep{Arnaud1996} v12.9.0 by adopting
$\chi^2$-statistics.

In this analysis, we make a reasonable assumption that the absorption
and the power-law photon index of Mrk~463E and Mrk~463W were constant
among the three observations of \textit{NuSTAR}, \textit{Chandra}, \textit{XMM-Newton} performed at
different epochs. We take into account possible time variability of the
AGN flux among the 3~epochs.
Utilizing the fact that 
the absorptions of the two AGNs are different from each other 
\citep{Bianchi2008} and that the low energy turnover observed in a few to 
$\sim$10 keV is very sensitive to column densities
of $N_{\rm H} \sim 10^{23}-10^{24}$ cm$^{-2}$, 
it is feasible to
{\it spectroscopically} separate their contributions in the \textit{NuSTAR} and
\textit{XMM-Newton} data even if they are spatially unresolved.

We take into account Galactic absorption in all spectral models,
fixing the hydrogen column density to 1.98 $\times 10^{20} {\rm
cm}^{-2}$, as estimated from the HI map of the region \citep{Kalberla2005}. To
correct for possible cross calibration uncertainties in absolute flux
among the different instruments, we multiply an energy-independent
constant factor (\textsf{const0}) to all the spectra by limiting its range to 0.8--1.2.
We set it to unity for the \textit{NuSTAR} and \textit{Chandra} spectra as calibration reference.

\subsection{Model I} \label{bozomath}

We first apply analytical models for the three regions 
(Mrk 463E, Mrk 463W, and Extended). 
The spectrum of each nucleus consists of five components.
In XSPEC terminology, it is expressed as:
\begin{align}
\textrm{Mrk\ } &\textrm{463E/W\ } \mathsf{= const0 * phabs} \notag\\
&\quad \mathsf{*(const1 * zphabs * cabs * zpowerlw * zhighect} \notag\\
&\quad \mathsf{+ const1 * zphabs * pexmon} \notag\\
&\quad \mathsf{+ const2 * zpowerlw * zhighect} \notag\\
&\quad \mathsf{+ apec + zgauss[\times4/1])}
\end{align}

The \textsf{const0} factor and \textsf{phabs} represent the cross-calibration
constant and the Galactic absorption, respectively.  The first term is the
transmitted component from the AGN.  It is modelled by an absorbed power
law with an exponential cutoff, which is fixed at 360~keV for
consistency with the \citet{Ikeda2009} model (see below)\footnote{We
utilize the \textsf{zhighect} model whose e-folding and cutoff energies are
fixed at 360 keV and 0.01 keV, respectively.}. The photon index and
normalization are tied together among all the spectra. The \textsf{const1}
factor accounts for time variability between different epochs, which is
fixed to unity for the \textit{NuSTAR} spectra but is left free to vary within 1/3 and 3 for the \textit{Chandra}
and \textit{XMM-Newton} spectra, consistently with the variability range 
observed from nearby AGNs over a timescale of $\sim$10 years 
\citep{Kawamuro2016}.
The \textsf{cabs} model takes into account the effect of Compton scattering in the
line-of-sight, which is however negligible when the column density is $N_{\rm H} < 10^{24}$~cm$^{-2}$.

The second term represents a reflection component from cold matter 
in the circumnuclear region, to which we refer to as the ``torus''.
We utilize the \textsf{pexmon} code \citep{Nandra2007}, which calculates
a reflected continuum with Fe and Ni 
fluorescence lines from optically-thick cold matter.
The photon index and the power-law normalization are linked to those of the
transmitted component.
The reflection strength, $R \equiv \Omega/2\pi$ ($\Omega$ is the solid
angle of the reflector), is set to be free within a physically 
plausible range of $0 < R \leq 2$. 
The inclination angle is fixed at 60$^\circ$ as a representative value.
We apply the same absorption to the reflection component as for the transmitted
component, which provides a better fit ($\Delta \chi^2 >2.46$) than the case
of no absorption.
Assuming that the reflection flux follows the transmitted one with 
a time delay of $<$a few years (i.e., its emitting region is located at
$\simlt$1 pc from the SMBH; \citealt{Minezaki2015,Gandhi2015}),
the same \textsf{const1} factor as to the first component is multiplied to 
this component. 
Assuming a constant absolute flux 
for the reflection component, we found values of the parameters
consistent within their 90\% uncertainty with those obtained 
allowing the constant to vary.

The third term represents a scattered component from the AGN, modelled
by an unabsorbed power law with an exponential cutoff at 360 keV. For
convenience, we link the power-law normalizations at 1 keV between the
transmitted and scattered components and multiply a constant factor
(\textsf{const2}) to the latter, which gives a scattering fraction, $f_{\rm
scat}$\footnote{
We note that contribution from high mass X-ray binaries
(HMXBs) in the host galaxy may be included in this unabsorbed power law component.}.
We do not link the photon indices of the transmitted and
scattered components for Mrk~463E, whereas we link them for Mrk~463W
because of the limited photon statistics of its \textit{Chandra} spectrum.
We assume that this component comes from regions larger than a
few pc scale and thus that absolute flux is constant over the 3 epochs.

The fourth term represents optically-thin thermal emission 
modeled by the \textsf{apec} code \citep{Smith2001},
which is expected if star forming activity is 
present in the host galaxy (see Section~\ref{sec4}).
The fifth term
represents additional emission line features modeled by \textsf{zgauss} that
cannot be explained by the lines included in the above \textsf{apec}
component; we consider that they are most likely 
recombination lines and 
radiative recombination continuum (RRC)
from a photoionized gas irradiated by the AGN \citep[e.g.,][]{Bianchi2006}.
We select them from the list of narrow lines summarized in Table~1 of
\citet{Bianchi2008} for each galaxy, and finally add them to the model
if they are found to significantly improve the fit with $\Delta \chi^2
>$ 2.71 (i.e., at a $>90\%$ confidence limit). Here we also consider
emission lines only detected with \textit{XMM-Newton}, assuming that they
originate predominantly from Mrk 463E, which is brighter than Mrk
463W and the extended emission.
We finally adopt lines at
0.654~keV (O~VIII K$\alpha$),  
0.739~keV (O~VII RRC), 1.342~keV (Mg~XI~K$\alpha$), and 
4.070~keV (Ca~XX~K$\alpha$) for Mrk~463E, and at 1.022~keV (Ne~X~K$\alpha$) for Mrk~463W.
We assume that the fluxes of the {\bf apec} and {\bf zgauss} components were constant over the 3 epochs.

The soft X-ray extended emission consists of three components,
expressed as:
\begin{align}
&\textrm{Extended\ } \mathsf{= const0 * phabs} \notag\\
&\qquad \mathsf{*(zpowerlw * zhighect + apec + zgauss)}. 
\end{align}
The first term represents a
scattered component from the AGN (Mrk 463E and/or Mrk 463W).
The second term represents 
optically-thin thermal emission (\textsf{apec}). The third term is an additional emission line
at 1.022 keV (Ne X K$\alpha$),
which is selected in the same way as for the Mrk 463E/W spectra.
We assume that the fluxes of all these components were constant over the 3 epochs.

\begin{figure*}
\plottwo{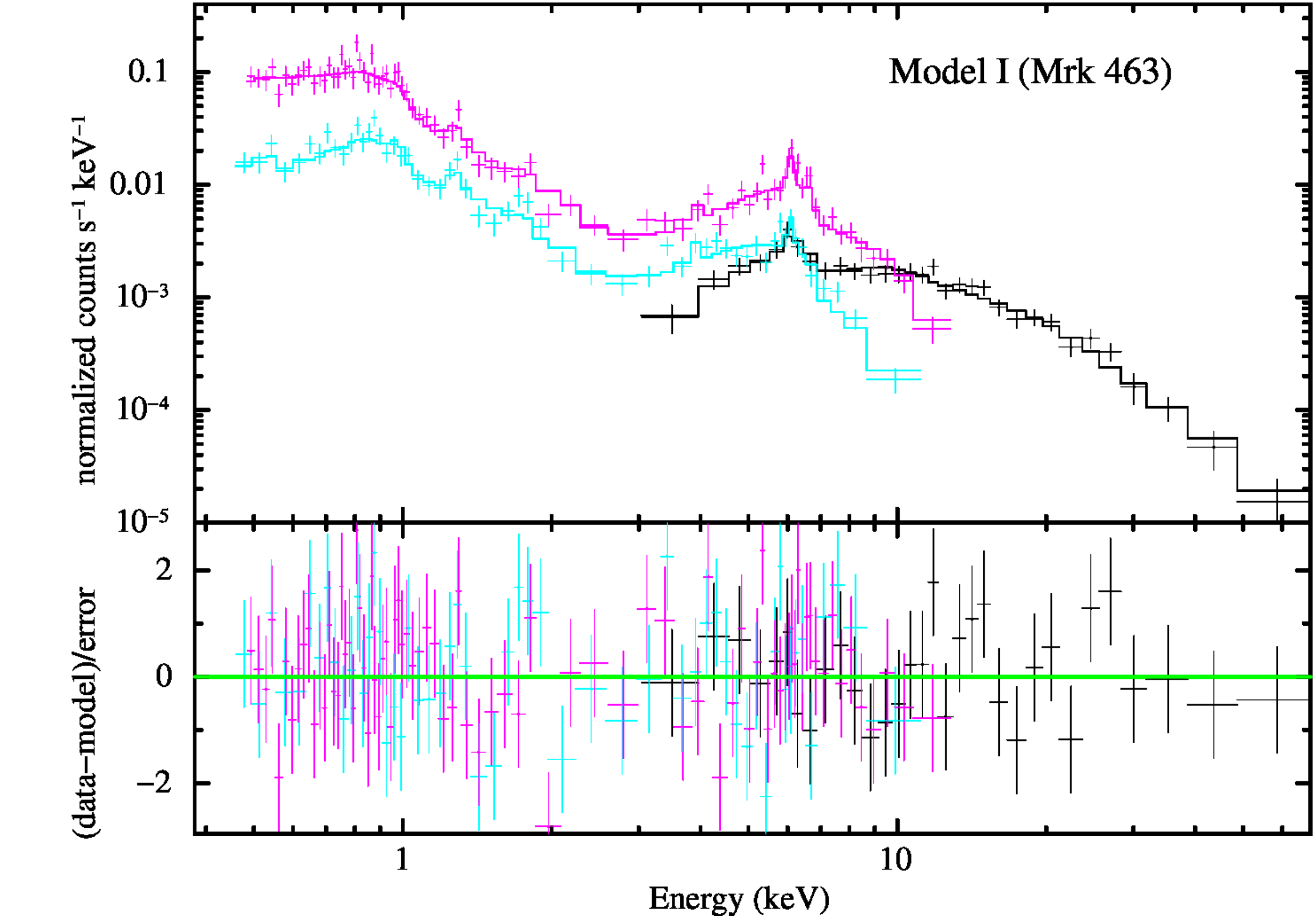}{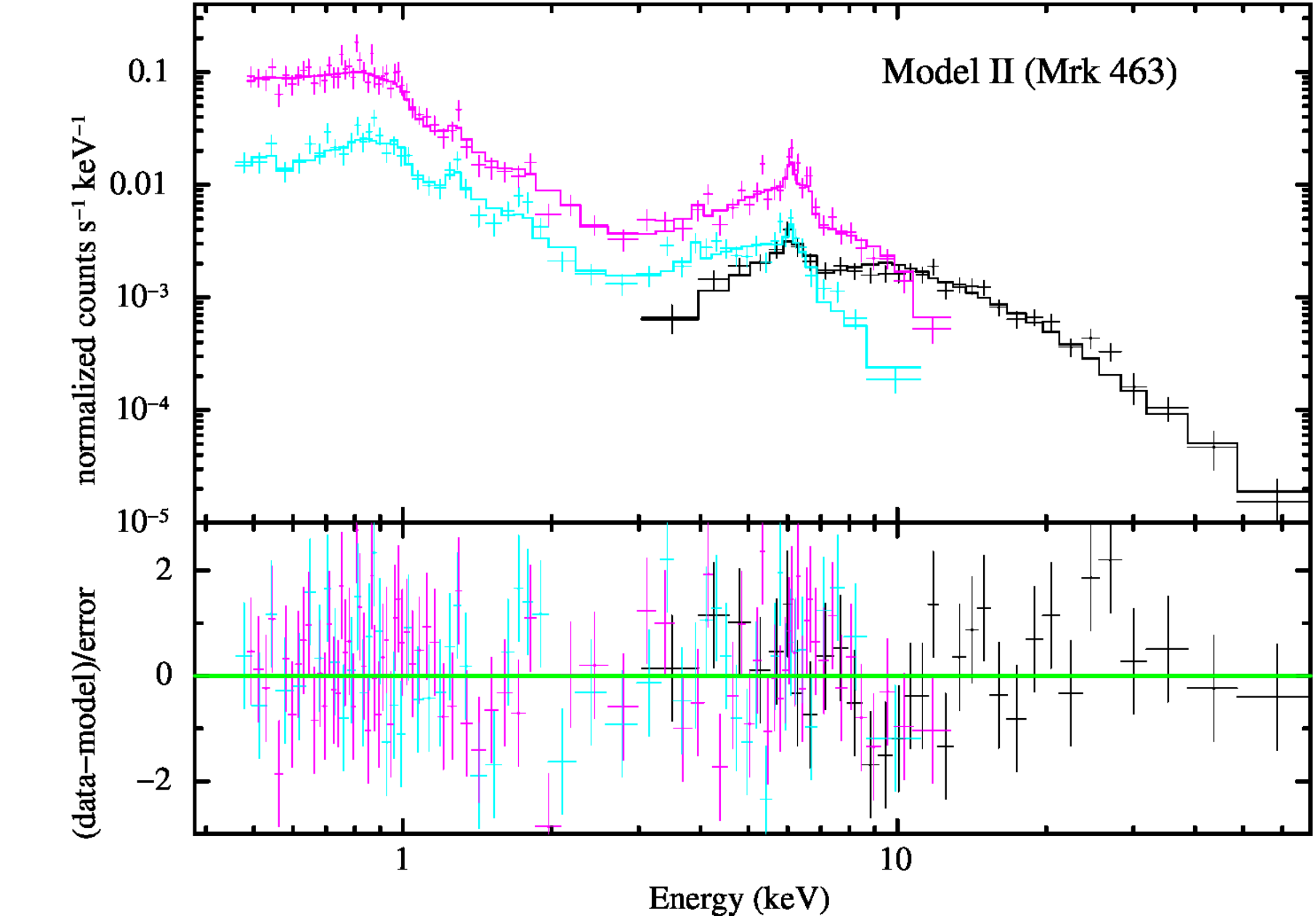}
\plottwo{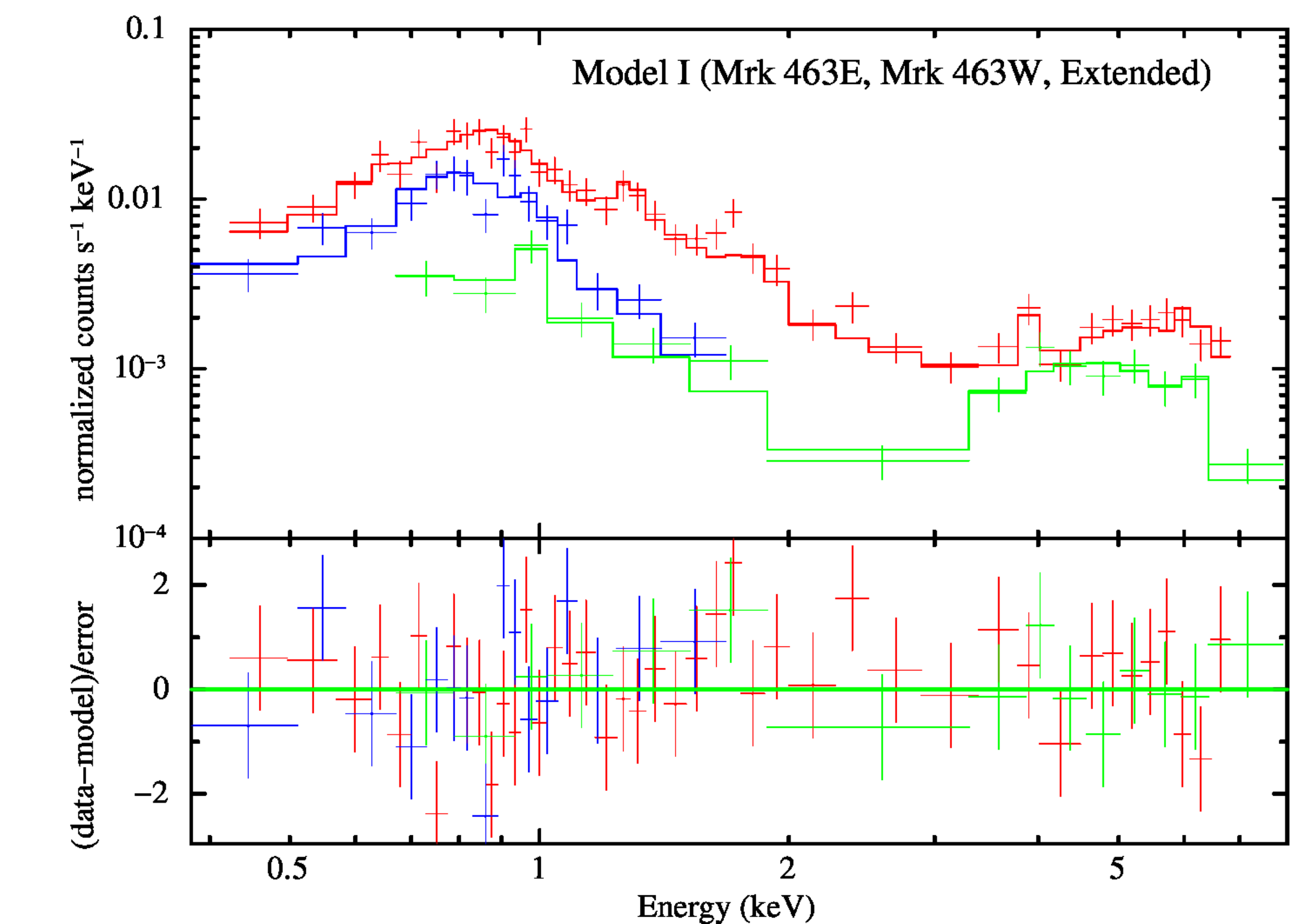}{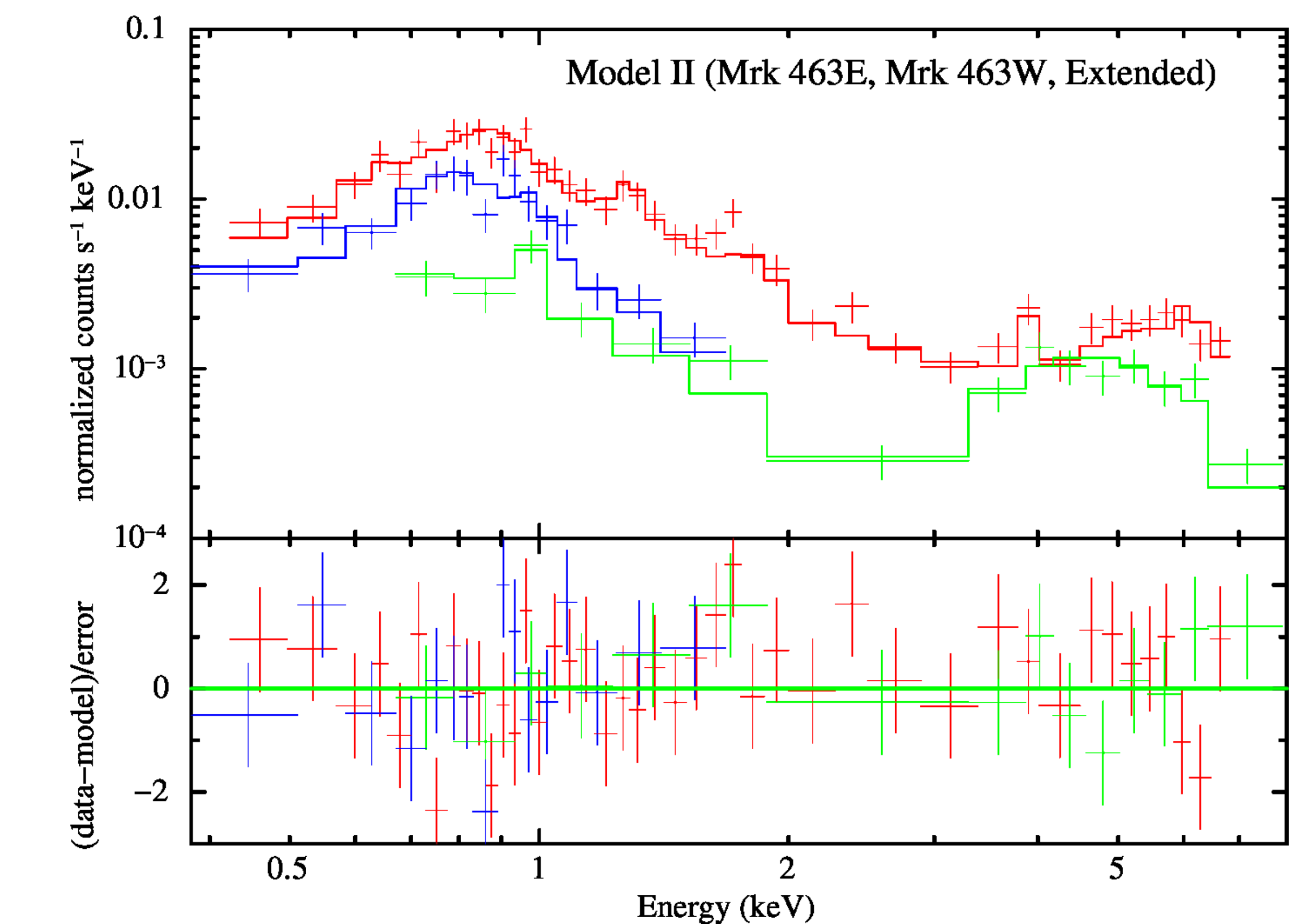}
\caption{
The observed spectra of Mrk 463
folded with the energy responses. 
The best-fit models are overplotted.
Left: Model~I. Right: Model~II. 
Upper: The black, light blue, and magenta crosses represent the \textit{NuSTAR}, \textit{XMM-Newton}/MOS, and \textit{XMM-Newton}/PN data of the total spectrum of Mrk 463, respectively.
Lower: The red, green, and blue crosses represent the \textit{Chandra} 
spectra of Mrk 463E, Mrk 463W, and the extended emission,
respectively. 
The residuals in units of $\sigma$ are plotted in the lower panels.
}
\label{fig2-fitting}
\end{figure*}

\begin{figure*}
\plottwo{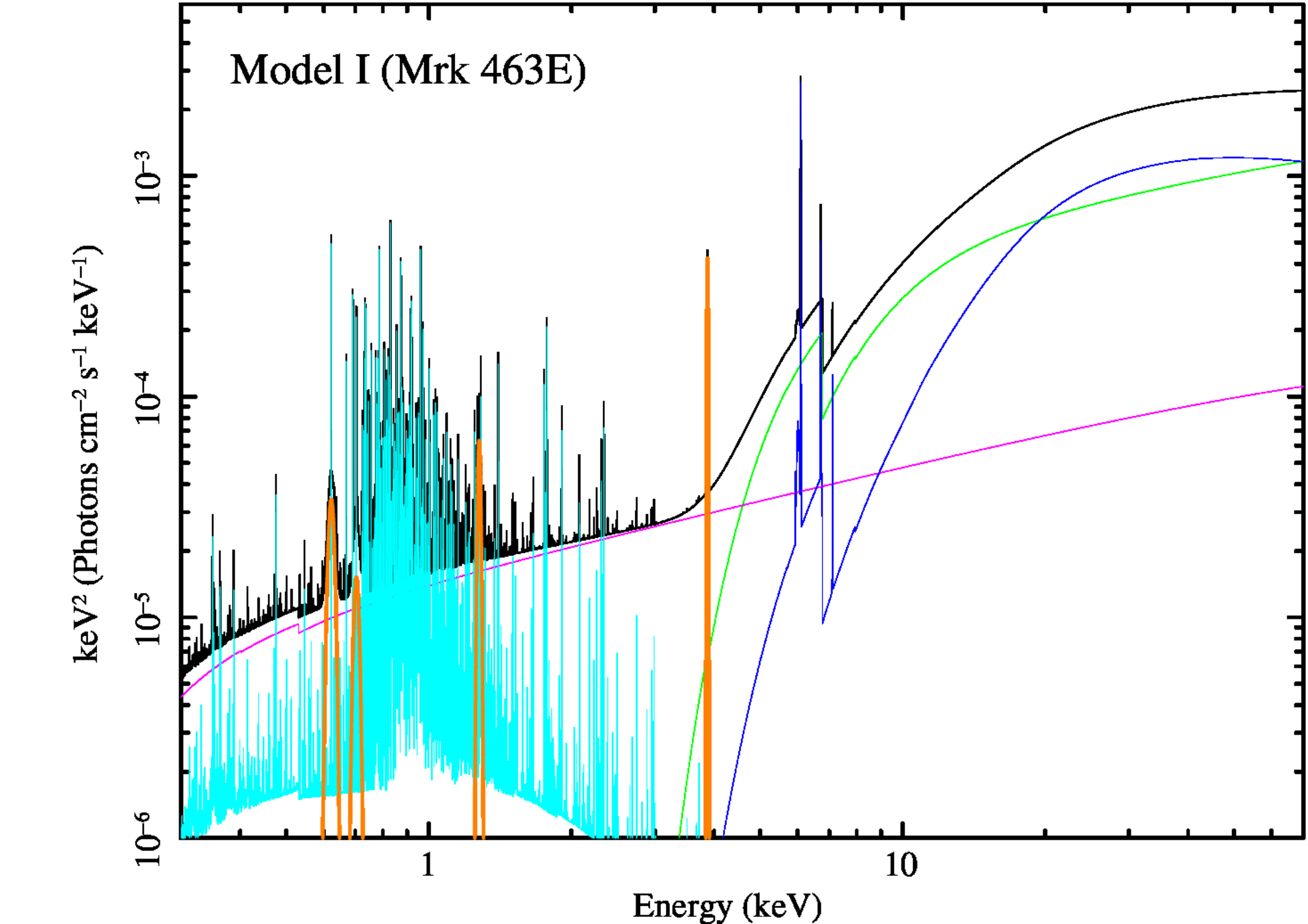}{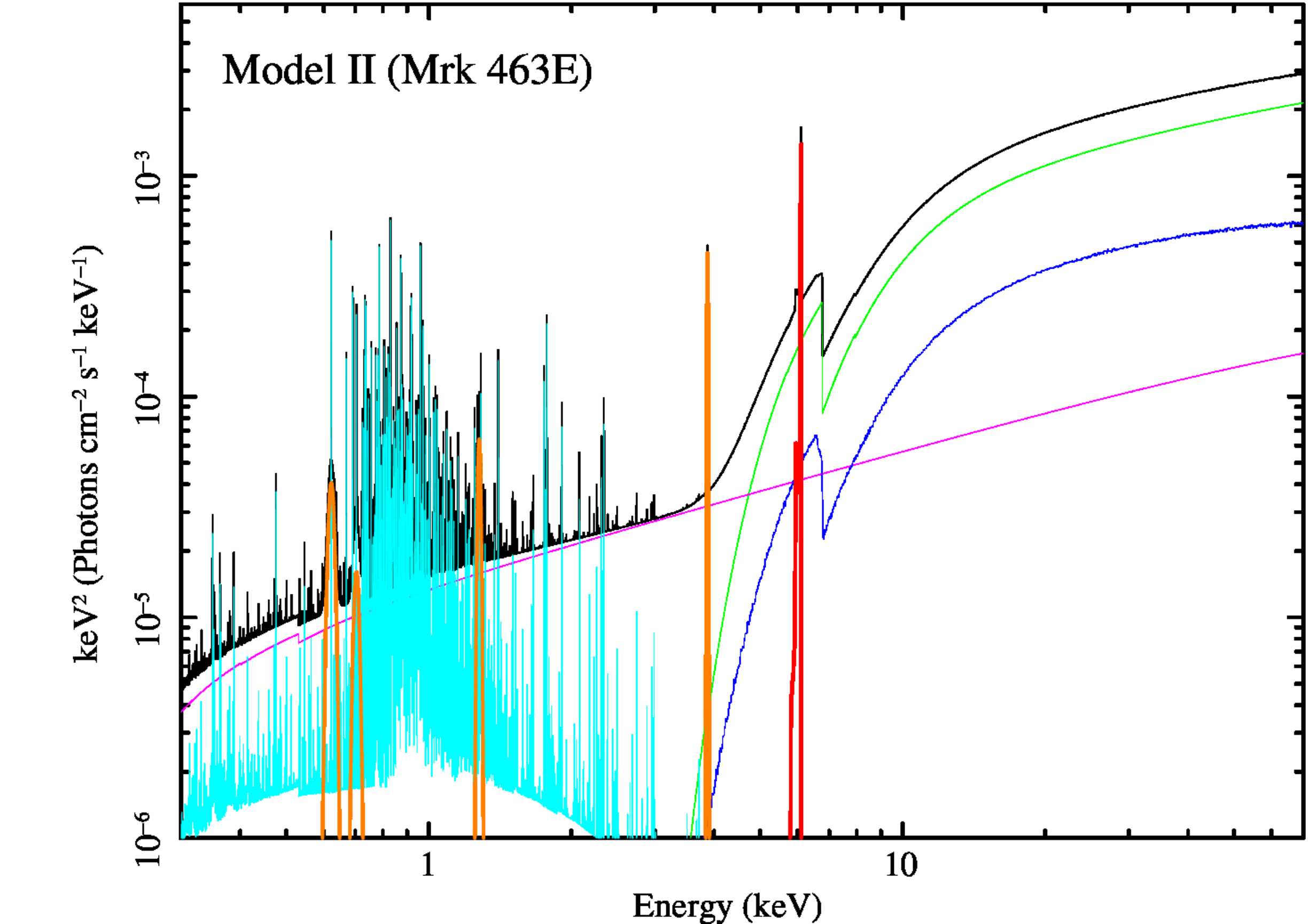}
\plottwo{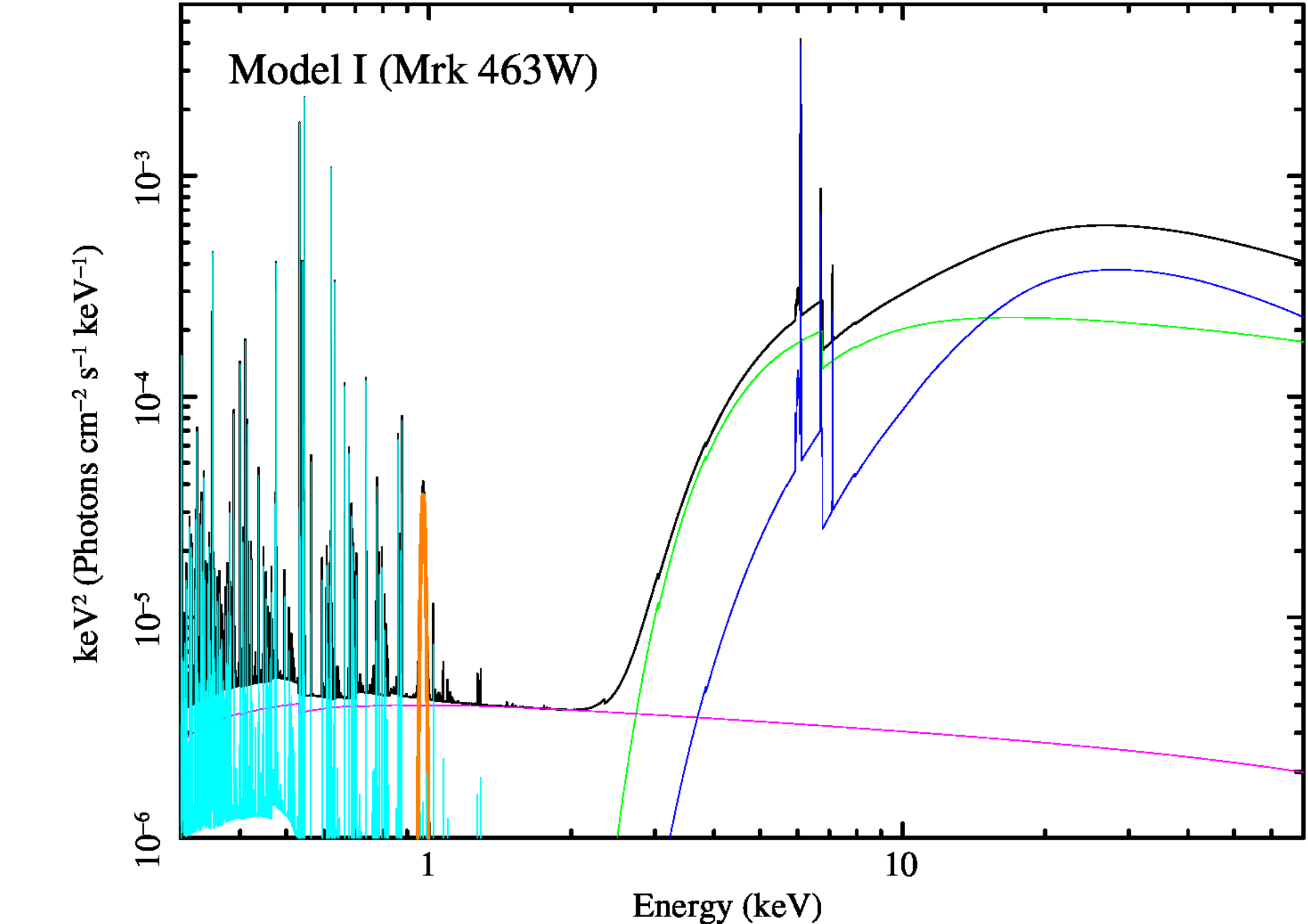}{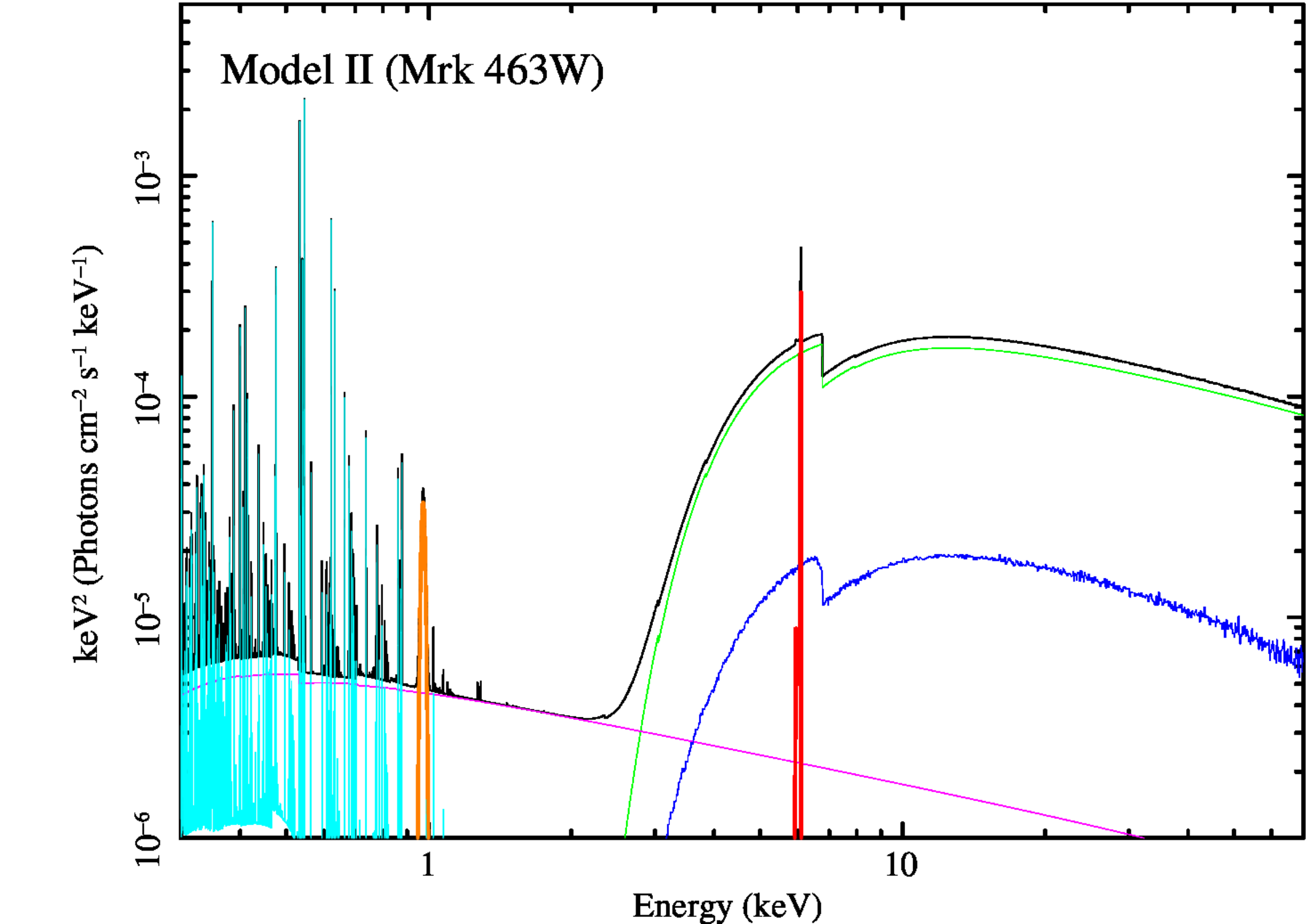}
\caption{The best-fit models in units of $E I_E$ (where $I_E$ is the
energy flux at the energy $E$). Left: Model I. Right: Model II. 
Upper: Mrk 463E. Lower: Mrk 463W. The black, green, magenta, blue, red, 
cyan, and orange lines represent the total model, transmitted component,
scattered component, reflection component, iron K$\alpha$ emission line,
optically-thin thermal emission, and additional emission lines,
respectively. 
}
\label{fig3-models}
\end{figure*}

Model I is found to well reproduce the overall spectra of Mrk 463
($\chi^2$/d.o.f = 239.9/215 for the total 6 spectra). 
The folded best-fit
models are overplotted in the left panels of Figure~\ref{fig2-fitting}, 
and the unfolded best-fit models for Mrk 463E and Mrk 463W are plotted
in the left panels of Figure~\ref{fig3-models}.
The best-fit parameters are
summarized in Table~\ref{tab2-parameters} along with the 
intrinsic AGN luminosities.
The hydrogen column density of Mrk 463E and Mrk 463W are estimated
to be $N_{\rm H} = 5.8^{+2.7}_{-1.4} \times 10^{23}$~cm$^{-2}$ and
$N_{\rm H} = 2.5\pm 0.6 \times 10^{23}$~cm$^{-2}$,
respectively.
The reflection components from the tori are significantly detected 
from both galaxies.

\subsection{Model II} \label{bozomath-2}

As a more physically self-consistent model, we next apply the
``e-torus'' model \citep{Ikeda2009} to the AGN spectra (Model~II). The
e-torus model is a Monte-Carlo based spectral model from a
uniform-density torus of nearly spherical geometry with bipolar cone
holes. The torus parameters are the ratio of the inner to outer radii
(fixed at 0.01), the half opening angle of the torus, the inclination
angle, and the column density along the equatorial plane. As the
incident spectrum, a power law with a photon index between 1.5--2.5 and
an exponential cutoff at 360 keV is assumed.  The reflected continuum
and iron-K fluorescence lines from the torus are provided as a table
model in XSPEC.

The spectrum of each nucleus is expressed in XSPEC terminology as:
\begin{align}
\textrm{Mrk\ } &\textrm{463E/W\ } \mathsf{= const0 * phabs} \notag\\
&\quad \mathsf{*(const1 * torusabs * zpowerlw * zhighect} \notag\\
&\quad \mathsf{+ const1 * zpowerlw * zhighect} \notag\\
&\qquad \mathsf{* mtable\{refl\_all\_20161115M.fits\}} \notag\\
&\qquad \mathsf{+ const1 * atable\{refl\_fe\_torus.fits\}} \notag\\
&\quad \mathsf{+ const2 * zpowerlw * zhighect} \notag\\
&\quad \mathsf{+ apec + zgauss[\times4/1])}
\end{align}

The model consists of five components:
(1) a transmitted component from the AGN, whose column density along the equatorial plane is self-consistently calculated from the torus 
geometry by \textsf{torusabs} \citep{Eguchi2011}, 
(2) a reflection component from the torus including an iron-K$\alpha$
emission line (6.4 keV), 
(3) a scattered component from the AGN, 
(4) optically-thin thermal emission in the host galaxy, and
(5) additional emission lines required from the data. 
The half opening angle and the inclination angle of the torus are fixed at
$60^{\circ}$ and
$80^{\circ}$\textcolor{red}{\footnote{In this case, the
ratio of the line-of-sight column density ($N_{\rm H}^{\rm LS}$ to
the column density along the equatorial plane ($N_{\rm H}^{\rm Eq}$) is 
$0.995$ \citep{Ikeda2009}.}}, 
respectively, since they are difficult to
constrain with our spectra, due to the limited photon statistics.
We have confirmed that the choice of these parameters 
does not affect significantly our results.

The spectrum of the soft X-ray extended emission is the same we used for Model I:
\begin{align}
&\textrm{Extended\ } \mathsf{= const0 * phabs} \notag\\
&\qquad \mathsf{*(zpowerlw * zhighect + apec + zgauss)}.
\end{align}

Model II also gives a good fit ($\chi^2$/d.o.f. = 257.9/217) of the overall
spectra of Mrk 463.
The best-fit models are plotted 
in the right panels of Figure~\ref{fig2-fitting} (folded spectra) 
and in the right panels of Figure~\ref{fig3-models} (unfolded spectra).
The best-fit parameters and intrinsic AGN luminosities 
are summarized in Table~\ref{tab2-parameters}.
The line-of-sight hydrogen column densities of Mrk~463E and Mrk~463W are
estimated to be $N_{\rm H} = 7.5^{+1.3}_{-1.2} \times 10^{23}$
cm$^{-2}$ and $N_{\rm H} = 2.9^{+0.6}_{-0.5} \times 10^{23}$
cm$^{-2}$, respectively.

\begin{deluxetable*}{llcccc}
\tablewidth{\textwidth}
\tablecaption{Best-fit Spectral Parameters \label{tab2-parameters}}

\tablehead{
\colhead{Region}           &
\colhead{No.}              &
\colhead{Parameter}        &
\colhead{Model I}  &
\colhead{Model II} &
\colhead{Units}            
}
\startdata
Mrk 463E & (1)  &$N_{\rm H}$               &$5.8^{+2.7}_{-1.4}$    &$7.5^{+1.3}_{-1.2}$    &$10^{23}$ cm$^{-2}$                      \\
         & (2)  &$\Gamma_{\rm AGN}$        &$1.5^{+1.1}_{-0.4}$    &$1.50^{+0.14}_{-0.00 (*)}$ &                                     \\
         & (3)  &$A_{\rm AGN}$             &$2.9^{+5.8}_{-2.2}$    &$5.5^{+3.0}_{-0.8}$    &$10^{-4}$ keV$^{-1}$ cm$^{-2}$ s$^{-1}$  \\
         & (4)  &$R$                       &$1.01^{+0.84}_{-0.86}$ &\nodata                &                                         \\
         & (5)  &$f_{\rm scat}$            &$5.5^{+18.8}_{-4.9}$   &$2.7^{+0.8}_{-1.1}$    &\%                                       \\
         & (6)  &$\Gamma_{\rm scat}$       &$1.47^{+0.22}_{-0.26}$ &$1.38^{+0.22}_{-0.27}$ &                                         \\
         & (7)  &$k$T                      &$0.89^{+0.07}_{-0.10}$ &$0.89^{+0.07}_{-0.10}$ &keV                                      \\
         & (8)  &$A_{\rm apec}$            &$0.96^{+0.21}_{-0.19}$ &$1.00^{+0.21}_{-0.19}$ &$10^{-5}$ cm$^{-5}$                      \\
         & (9)  &$A$(0.654 keV)            &$2.5\pm 1.9$           &$3.0\pm 1.9$           &$10^{-6}$                                \\
         & (10) &$A$(0.739 keV)            &$0.9^{+1.6}_{-0.9 (*)}$&$0.9^{+1.6}_{-0.9 (*)}$&$10^{-6}$                                \\
         & (11) &$A$(1.342 keV)            &$1.0\pm 0.5$           &$1.0\pm 0.5$           &$10^{-6}$                                \\
         & (12) &$A$(4.070 keV)            &$0.7\pm 0.4$           &$0.8\pm 0.4$           &$10^{-6}$                                \\
         & (13) &$N_{\rm Chandra}$         &$1.7^{+1.3 (*)}_{-0.8}$&$1.33^{+0.45}_{-0.35}$ &                                         \\
         & (14) &$N_{\rm XMM}$             &$1.6^{+1.4 (*)}_{-0.5}$&$1.26^{+0.31}_{-0.26}$ &                                         \\
         & (15) &$F_{\rm 2-10 keV}$        &$3.11\pm 0.11$         &$3.98\pm 0.13$         &$10^{-13}$ erg cm$^{-2}$ s$^{-1}$        \\
         & (16) &$L_{\rm 2-10 keV}$        &$0.85^{+0.41}_{-0.29}$ &$1.64\pm 0.21$         &$10^{43}$ erg s$^{-1}$                   \\
\hline

Mrk 463W & (17) &$N_{\rm H}$               &$2.5\pm 0.6$           &$2.9^{+0.6}_{-0.5}$    &$10^{23}$ cm$^{-2}$                      \\
         & (18) &$\Gamma_{\rm AGN}$        &$2.1^{+1.0}_{-0.7}$    &$2.42^{+0.08 (*)}_{-0.45}$ &                                     \\
         & (19) &$A_{\rm AGN}$             &$4.9^{+18}_{-3.7}$     &$8.2^{+5.7}_{-5.3}$    &$10^{-4}$ keV$^{-1}$ cm$^{-2}$ s$^{-1}$  \\
         & (20) &$R$                       &$2.0^{+0.0 (*)}_{-1.2}$&\nodata                &                                         \\
         & (21) &$f_{\rm scat}$            &$1.0^{+1.6}_{-0.6}$    &$0.7^{+0.9}_{-0.3}$    &\%                                       \\
         & (22) &$k$T                      &$0.16^{+0.04}_{-0.12}$ &$0.15^{+0.04}_{-0.11}$ &keV                                      \\
         & (23) &$A_{\rm apec}$\tablenotemark{\dag}&$1.30_{-0.57}$ &$1.50_{-0.76}$         &$10^{-5}$ cm$^{-5}$                      \\
         & (24) &$A$(1.022 keV)            &$1.0\pm 0.6$           &$0.9\pm 0.6$           &$10^{-6}$                                \\
         & (25) &$N_{\rm Chandra}$         &$0.64^{+0.50}_{-0.20}$ &$0.82^{+0.65}_{-0.24}$ &                                         \\
         & (26) &$N_{\rm XMM}$             &$0.60^{+0.39}_{-0.27 (*)}$ &$0.79^{+0.51}_{-0.26}$ &                                     \\
         & (27) &$F_{\rm 2-10 keV}$        &$3.24\pm 0.23$         &$2.31\pm 0.17$         &$10^{-13}$ erg cm$^{-2}$ s$^{-1}$        \\
         & (28) &$L_{\rm 2-10 keV}$        &$0.56^{+0.22}_{-0.20}$ &$0.62^{+0.29}_{-0.28}$ &$10^{43}$ erg s$^{-1}$                   \\
\hline
Extended & (29) &$\Gamma_{\rm Ext}$        &$2.58^{+0.48}_{-0.44}$ &$2.50^{+0.00 (*)}_{-0.36}$ &                                     \\
         & (30) &$A_{\rm Ext}$             &$5.1\pm 1.5$           &$5.2\pm 1.4$           &$10^{-6}$ keV$^{-1}$ cm$^{-2}$ s$^{-1}$  \\
         & (31) &$k$T                      &$0.66^{+0.12}_{-0.08}$ &$0.65^{+0.11}_{-0.08}$ &keV                                      \\
         & (32) &$A_{\rm apec}$            &$0.53\pm 0.10$         &$0.53\pm 0.10$         &$10^{-5}$cm$^{-5}$                       \\
         & (33) &$A$(1.022 keV)            &$1.1\pm 0.7$           &$1.2\pm 0.7$           &$10^{-6}$                                \\
         & (34) &$F_{\rm 0.5-2 keV}$       &$2.43\pm 0.19$         &$2.45\pm 0.19$         &$10^{-14}$erg cm$^{-2}$ s$^{-1}$         \\
         & (35) &$L_{\rm 0.5-2 keV}$       &$1.49\pm 0.11$         &$1.49\pm 0.11$         &$10^{41}$erg s$^{-1}$                    \\
\hline
         & (36) &$C_{\rm XMM-MOS}$         &$1.15^{+0.05 (*)}_{-0.08}$ &$1.15^{+0.05 (*)}_{-0.07}$ &                                 \\
         & (37) &$C_{\rm XMM-PN}$          &$1.10\pm 0.07$         &$1.09\pm 0.07$         &                                         \\
         &      &$\chi^2/d.o.f.$ $(\chi^2_r)$  &239.9/215 (1.12)   &257.9/217 (1.19)
\enddata

\tablecomments
{
(1) Hydrogen column density.
(2) Power law photon index of the transmitted component.
(3) Power law normalization at 1 keV.
(4) Reflection strength ($R = \Omega/2\pi$) in the \textsf{pexmon} model.
(5) Scattering fraction.
(6) Power law photon index of the scattering component.
(7) Temperature of the \textsf{apec} model.
(8) Normalization of the \textsf{apec} model.
(9) Normalization of the O ${\rm VIII}$ K$\alpha$.
(10) Normalization of the O ${\rm VII}$ RRC.
(11) Normalization of the Mg ${\rm XI}$ K$\alpha$ line.
(12) Normalization of the Ca ${\rm XX}$ K$\alpha$ line.
(13) Time variability constant of \textit{Chandra} relative to \textit{NuSTAR}.
(14) Time variability constant of \textit{XMM-Newton} relative to \textit{NuSTAR}.
(15) Observed flux in the 2--10 keV band (at the \textit{NuSTAR} observation). 
The error is estimated by 
fixing all the spectral parameters at the best-fit values 
except the overall normalizations for the three regions.
(16) Intrinsic (de-absorbed) AGN luminosity in the rest frame 2--10 keV
 band (at the \textit{NuSTAR} observation).
The error is estimated by fixing the photon index at the best-fit value.
(17)--(23) Parameters for Mrk 463W corresponding to (1)--(5) and (7)--(8), respectively.
(24) Normalization of the Ne X K$\alpha$ line.
(25)--(28) Parameters for Mrk 463W corresponding to (13)--(16), respectively.
(29) Power-law photon index.
(30) Power-law normalization 1 keV.
(31) Temperature of the \textsf{apec} model.
(32) Normalization of the \textsf{apec} model.
(33) Normalization of the Ne {\rm X} K$\alpha$ line.
(34) Observed flux of the extended emission in the 0.5--2 keV band.
(35) Luminosity of the extended emission in the rest-frame 0.5--2 keV
band. 
The same relative error as in the observed flux is attached.
(36) Cross-normalization of \textit{XMM}/MOSs relative to \textit{NuSTAR}.
(37) Cross-normalization of \textit{XMM}/PN relative to \textit{NuSTAR}.
}
\tablenotetext{$\dagger$}{The upper limit cannot be well constrained owing to the lack of data below 0.7 keV in the Chandra Mrk 463W spectrum.}
\tablenotetext{*}{The parameter reaches a limit of its allowed range.}
\end{deluxetable*}

\section{Discussion}
\label{sec4}

We have presented our broadband X-ray (0.4--70 keV) spectral analysis of
Mrk 463 by utilizing all available data of \textit{NuSTAR}, \textit{Chandra}, and
\textit{XMM-Newton} observations, carried out in 2014, 2004, and 2001, respectively. 
With a help
of the \textit{Chandra} data, we are able to separate broadband spectra of Mrk
463E, Mrk 463W, and of the extended emission, by 
assuming that their spectral shapes (not fluxes) were constant among
these epochs. In this section, we discuss our results, focusing on the two
nuclei, Mrk 463E and Mrk 463W.

We have shown that both the analytical model (Model~I) and the numerical
torus model (Model II) well reproduce the spectra of Mrk 463E and Mrk 463W.
\citet{Bianchi2008} adopted an empirical model consisting of an absorbed
power-law, unabsorbed power-law, and individual emission lines, whereas
we include a reflection component from the torus and an
optically-thin thermal emission component as physically motivated
models. 
Although it is difficult to disentangle origins in a
collisional gas or a photoionized gas for the soft X-ray emission
using the CCD data alone, 
we verify that the derived soft X-ray
luminosities of the thermal components are reasonable as compared with the
star formation rate of the galaxies.
From the 42.5--122.5 $\mu$m luminosity ($5.7 \times 10^{44}$ erg s$^{-1}$ as estimate from the IRAS 60 $\mu$m and 100 $\mu$m photometries) 
and the 8--1000 $\mu$m luminosity ($2.4 \times 10^{45}$ erg s$^{-1}$, \citealt{Imanishi2014}), of Mrk 463,
the 0.5--2 keV luminosity 
originating from the star formation activity
is expected to be $1.3\times 10^{41}$ erg s$^{-1}$ and $7 \times
10^{40}$ erg s$^{-1}$ (mean values), 
based on the relations of \citet{Ranalli2003} and
\citet{Iwasawa2011}, respectively.
The total observed 0.5--2 keV
luminosity of the \textsf{apec} components in Mrk 463 
($2.9 \times 10^{41}$ erg s$^{-1}$) is 
consistent with these predictions within the scatter.
With Model~I, where the
\textsf{pexmon} code is utilized, we detect significant reflection
components
from both galaxies, as expected from the presence of strong
iron-K$\alpha$ emission lines in the \textit{Chandra} and
\textit{XMM-Newton} spectra.  These reflection features are consistently
explained by the numerical torus model (Model~II).  We confirm that the
basic spectral parameters obtained with Model~I and Model~II are
consistent with each other. Hereafter, we refer to the results obtained
with Model~II unless otherwise stated.

\subsection{Mrk 463E}

We have determined the line-of-sight column density of Mrk 463E to be
7.5 $\times 10^{23}$ cm$^{-2}$, which is consistent with the result by
\citet{Bianchi2008}, and thus confirm that the AGN is heavily obscured
but not Compton-thick.  We obtain a flatter slope ($\Gamma \simeq 1.5$)
for the AGN transmitted component than \citet{Bianchi2008} ($\Gamma
\simeq 2.3$). This is most likely because an optically-thin thermal
component was not included in their model, leading to a steeper slope of
the unabsorbed power law component linked to that of the transmitted
component. 
Our analysis suggests that the fluxes of
Mrk 463E in the \textit{NuSTAR} observation in 2014 were $N_{\rm
Chandra} \approx 1.33$ and $N_{\rm XMM} \approx 1.26$~times lower than
those in the \textit{Chandra} (2004) and \textit{XMM-Newton} (2001)
observations, respectively.  The best-fit intrinsic (de-absorbed)
2--10~keV luminosities range from 1.6 $\times 10^{43}$ erg s$^{-1}$
(2014) to 2.2 $\times 10^{43}$ erg s$^{-1}$ (2004) , confirming that Mrk
463E has a Seyfert class AGN.  In the following discussions centered on
the X-ray luminosity, we take into account the range of this
variability.

Comparison of luminosities between X-rays and high-excitation mid-IR
lines from the narrow line region (NLR), such as [O IV] 25.89 $\mu$m and [Ne V] 14.32 $\mu$m,
gives useful insight into the spectral energy distribution (SED) and
circumnuclear structure of the AGN.
Since such mid-IR lines emitted from
the NLR are less attenuated by dust in the host galaxy and 
are less contaminated by star-formation activities, they are generally thought
to be better indicators of the intrinsic AGN luminosity
than the widely used [O III] $\rm \lambda$5007 line \citep{Heckman2005}.
Since these MIR
lines are produced by irradiation of UV photons from the AGN, their
luminosity ratios to X-rays reflect the intrinsic AGN SED in the UV to
X-ray band, unless the nucleus is deeply ``buried'' by dust (i.e., the obscurer has 
a large covering factor) and hence the NLR is underdeveloped \citep{Oda2017}.

\begin{figure}
    \centering
    \includegraphics[keepaspectratio,scale=0.7]
    {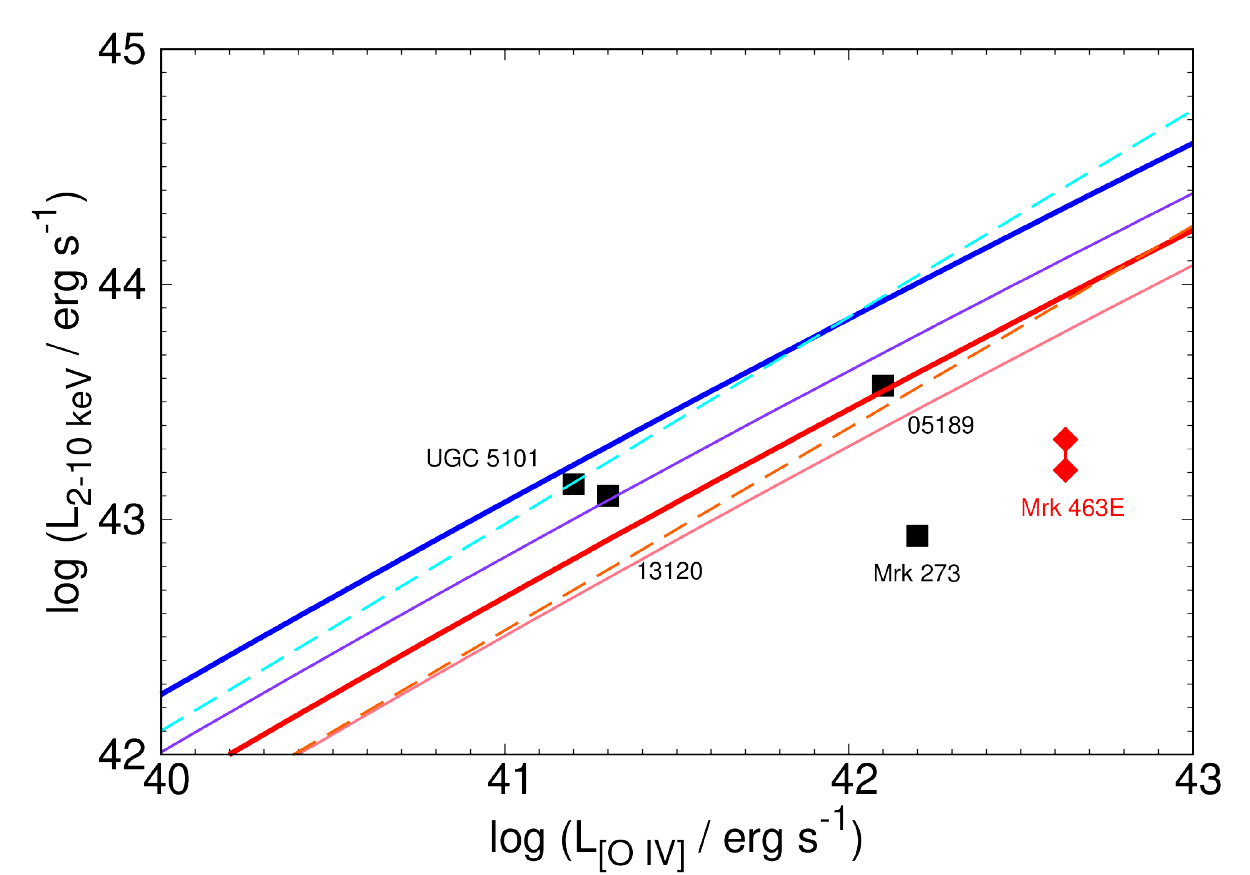}
    \caption{
The 2-10 keV versus  [O IV] 25.89 $\mu$m luminosity plot of local 
U/LIRGs. Mrk 463E is marked with the red diamonds
referring to the X-ray luminosities in 2014 (lower) and 2004 (upper).
The Compton-thick ULIRGs in \citet{Teng2015} and UGC 5101 in \citep{Oda2017} are plotted with the black squares.
The bold blue/red curves and solid purple/pink lines represent the averaged relations for local Seyfert 1s/2s and Compton-thin/thick Seyfert 2s \citep{Rigby2009}, respectively.
The dashed cyan/orange lines show the Seyfert 1s/2s relations obtained by \citet{Liu2014}.
    }
    \label{fig4-oiv-x}
\end{figure}

Figure \ref{fig4-oiv-x} plots the relation between the de-absorbed 2--10 keV luminosity
and the [O IV] 25.89 $\mu$m luminosity for Mrk 463E \citep{Armus2004}.
As noticed, the luminosity ratio between X-ray (2-10 keV)
and [O IV] 25.89 $\mu$m of Mrk 463E is smaller than the averaged values of
normal Seyfert galaxies by a factor of $\sim$5, indicating
that it is ``X-ray weak'' relative to the UV luminosity.
The X-ray underluminosity was also implied from the broad optical/NIR
emission lines \citep{Imanishi2004} and from the [O III] 
$\rm \lambda$5007 line \citep{Bianchi2008}.
The X-ray faint SED (i.e., a large bolometric to X-ray luminosity ratio)
implies that the AGN has a high Eddington ratio
\citep{Vasudevan2007}, that is, contain a rapidly growing SMBH.
We also note that this result provides no evidence that 
the nucleus is deeply ``buried'' by dust, because it would lead to the 
opposite trend (i.e., brighter in the X-rays than in [O IV]).
For reference, we also plot the data of IRAS 05189-2524, IRAS 13120-5453, Mrk
273, and UGC 5101 \citep{Veilleux2009,Teng2015,Oda2017}, 
which are ULIRGs 
observed with \textit{NuSTAR}. 
We note that Mrk 273, a late-stage merging galaxy with
one or two obscured AGNs \citep{Iwasawa2011-Mrk273}, 
may be a similar system to Mrk 463, 
whereas the other ULIRGs showing same
X-ray vs [O IV] relations as normal Seyferts may contain 
``buried'' AGNs with high Eddington ratios, where the [O IV] emission is
significantly suppressed \citep{Oda2017}.

We quantitatively estimate the Eddington ratio of Mrk~463E, using
$L^{\prime}$-band (3.8 $\mu$m), [O IV] 25.89 $\mu$m, and [Ne~V] 14.32~$\mu$m luminosities.
The $L^{\prime}$-band (3.8~$\mu$m) emission is
produced by AGN-heated hot dust in a vicinity ($\sim$ a few pc) of an AGN
\citep{Barvainis1987}, and hence its luminosity (after subtracting
a starburst contribution) is also a good indicator of the AGN bolometric
luminosity. By comparing the $K$-band (2.2 $\mu$m) and $L^{\prime}$-band
luminosities, \citet{Imanishi2014} estimated the AGN contribution to the
$L^{\prime}$-band luminosity of Mrk 463E is $\sim$100\%. Table \ref{tab3-bol-lumin}
summarizes the luminosities in the $L^{\prime}$-band (3.8 $\mu$m), [O IV]
25.89 $\mu$m, and [Ne V] 14.32~$\mu$m, as well as the corresponding
bolometric luminosities based on the relations by \citet{Risaliti2010}, \citet{Rigby2009}, and \citet{Satyapal2007},
respectively. 
These bolometric correction factors depend on the torus opening angle;
here we assume that those of Mrk 463E is similar to the averaged values
of normal Seyfert galaxies because its nucleus is unlikely to be deeply
buried (see above). We find these bolometric luminosities agree with one
another within a factor of 2. They are also similar to the total
infrared luminosity of Mrk~463 (sum of Mrk~463E and Mrk~463W),
indicating that Mrk 463E is an AGN-dominated LIRG. Adopting the black
hole mass of Mrk 463E, 5.5 $\times 10^7 M_{\odot}$, as determined by \citet{Dasyra2011} using the $M_{\rm BH} - \sigma_{*}$ relation, we calculate the Eddington ratio
($\lambda_{\rm Edd}$) for each luminosity estimate (Table~3).  We find
$\lambda_{\rm Edd} = 0.4-0.8$, thus confirming that the AGN in Mrk 463E
has a high Eddington ratio. We also confirm that the X-ray to
bolometric luminosity ratios is large, $\kappa_{\rm 2-10 keV} \equiv
L_{\rm bol}/L_{\rm 2-10 keV} \sim$ 110--410
(Table~\ref{tab3-bol-lumin}).

\begin{deluxetable*}{ccccc}
\tablewidth{\textwidth}
\tablecaption{Summary of Bolometric Luminosity and Eddington Ratio of Mrk 463E
\label{tab3-bol-lumin}}
\tablehead{
\colhead{Method}  &
\colhead{$L_{\rm obs}$}  &
\colhead{$L_{\rm bol}^{\rm AGN}$}  &
\colhead{$\kappa_{\rm 2-10 keV}$}  &
\colhead{$\lambda_{\rm Edd}$}  
\\
(1)&(2)&(3)&(4)&(5)
}
\startdata
$L^{\prime}$-band (3.8 $\mu$m) & 550  & 2.75 & 110--190 & 0.4 \\
$[$O IV$]$ (25.89 $\mu$m)      & 4.27 & 5.63 & 230-390 & 0.8 \\
$[$Ne V$]$ (14.32 $\mu$m)      & 1.15 & 5.88 & 240-410 & 0.8 
\enddata
\tablecomments
{
(1) AGN bolometric luminosity indicator.
(2) Its observed luminosity in units of $10^{42}$ erg s$^{-1}$.
(3) Estimated AGN bolometric luminosity in units of $10^{45}$ erg s$^{-1}$.
(4) X-ray (2--10 keV) to bolometric correction factor. 
The statistical error in the X-ray luminosity and time
variability range 
based on Model~II are taken into account.
(5) Eddington ratio calculated with the SMBH mass by \citet{Dasyra2011}.
}
\end{deluxetable*}

We note that AGN time variability 
(i.e., the X-ray observations were performed by chance
when the AGN activity was largely suppressed) is unlikely to
be a major origin of the X-ray weakness. The $L^{\prime}$-band emission
predominantly comes from AGN-heated dust of $\sim$1000 K, whose
sublimation radius is estimated to be $\approx$2--3 pc for $L_{\rm bol}
= (3-6)\times10^{45}$ erg s$^{-1}$ \citep{Barvainis1987}. Thus, the
$L^{\prime}$-band luminosity of this object should reflect its past AGN
activity over several to ten years. Since the $L^{\prime}$-band data
of Mrk 463E were taken in 2012 \citep{Imanishi2014}, they can be reasonably
compared with the X-ray luminosities measured in 2004 (with \textit{Chandra}) and 2001 (with \textit{XMM-Newton}), on which our discussion
is based.

To summarize, we confirm that the AGN in Mrk 463E has a high Eddington
rate (i.e., the SMBH is rapidly growing) and is X-ray underluminous
compared with normal Seyferts. These properties are consistent with the
general trend between $\lambda_{\rm Edd}$ and $\kappa_{\rm 2-10 keV}$
\citep{Vasudevan2007}. We infer that the merger triggered efficient gas fueling
onto the SMBH. Unlike typical ULIRGs, however, the nucleus of Mrk 463E
is not deeply buried by dust. This is in line with the evolutionary
scenario of mergers in which obscuration of the nuclei by galactic-scale
gas/dust becomes more significant toward later stages \citep{Ricci2017}.

\subsection{Mrk 463W}

We determine the absorption of Mrk 463W to be 2.9 $\times 10^{23}$
cm$^{-2}$, confirming that the galaxy contains a Compton-thin obscured
AGN. The column density and photon index ($\Gamma\simeq2.4$) are
consistent with the results by \citet{Bianchi2008}. Our analysis
suggests that the flux of Mrk 463E in 2014 was higher than those in 2004
and 2001 by factors of $N_{\rm Chandra} \approx 0.82$ and $N_{\rm XMM} \approx
0.79$, respectively, although this variation is not statistically
significant. The best-fit intrinsic 2--10 keV luminosities range from
6.2 $\times 10^{42}$~erg~s$^{-1}$ (2014) to 5.1 $\times 10^{42}$~erg~s$^{-1}$ (2004).  Below we discuss our results by comparison with the
$L^{\prime}$-band luminosity. Unfortunately, neither the [O~IV] 25.89~$\mu$m and [Ne V] 14.32 $\mu$m luminosities nor the black hole mass are
not available for Mrk 463W.

We obtain a bolometric luminosity of Mrk 463W of $8.5 \times 10^{42}$
erg s$^{-1}$, using the $L^{\prime}$-band luminosity after subtracting
the 68\% starburst contribution \citep{Risaliti2010,Imanishi2014}. This
yields an unusually small bolometric to 2--10 keV luminosity ratio,
$\kappa_{\rm 2-10 keV} \sim 1-3$, as compared with typical AGNs
($\kappa_{\rm 2-10 keV} \sim$ 10--70; \citealt{Vasudevan2007}). 
We consider three possibilities to explain this very small value. The
first possibility is that the AGN torus in Mrk 463W has a very low
covering factor,
to which the mid-IR luminosity is approximately proportional.
We infer it unlikely because this
contradicts with the strong reflection component 
($R > 0.8$) seen in the
X-ray spectrum. 
The second one is time variability. When an AGN is
rapidly brightening from a quiescent state, it becomes bright in the
UV/X-ray band first before the radiation reaches to circumnuclear dust
that emit mid-IR radiation. 
The fact that Mrk 463W was already X-ray bright in 2001 but faint in the
$L^{\prime}$-band in 2012 \citep{Imanishi2014} suggests that the
$L^{\prime}$-band emission would be mainly produced at distances of $>$3 pc.
These distances are, however, much larger than the sublimation radius of
dust with $\sim$1000 K for $L_{\rm bol}$
$\sim 10^{44}$ erg s$^{-1}$ (converted from the X-ray luminosity with
$\kappa_{\rm 2-10 keV} \sim$20 by assuming a normal SED.), 
which is $\sim$ 0.4 pc \citep{Barvainis1987}.
Hence, we infer that time variability is unlikely to be a main
cause of the small $\kappa_{\rm 2-10 keV}$ value. The third
possibility, the most plausible one, is that the AGN has a low Eddington
ratio ($<10^{-3}$), in which case the X-ray to UV luminosity ratio
becomes large \citep{Ho2008}. In fact, this could be the case if the
SMBH mass in Mrk 463W is similar to or larger than that in Mrk 463E,
considering that the estimated bolometric luminosity of Mrk 463W is
$\sim$300 times smaller than that of Mrk 463E. This can be directly
tested by future measurements of the SMBH mass of Mrk 463W. We speculate
that this object might be still in an early phase of AGN activity
triggered by the merger.

\section{Conclusion}
\label{sec5}

We have analyzed the broadband spectra (0.4--70 keV) of the
double-nucleus LIRG Mrk 463 for the first time, observed with
\textit{NuSTAR}, \textit{Chandra}, and \textit{XMM-Newton}. An
analytical model utilizing the \textsf{pexmon} reflection code and a
numerical model utilizing the e-torus model well reproduce the
overall spectra. We have determined the absorption column densities of
Mrk 463E and Mrk 463W to be 7.5 $\times 10^{23}$ cm$^{-2}$ and 2.9 $\times 10^{23}$ cm$^{-2}$, respectively. Significant
reflection components are detected in both AGNs, suggesting that their
tori are well developed.

The luminosity ratio between X-ray (2--10 keV) and [O~IV] 25.89 $\mu$m
of Mrk 463E is smaller than those of normal Seyferts by a factor of
$\sim 5$, indicating that the intrinsic AGN SED is X-ray weak with
respect to the UV luminosity. In fact, the bolometric luminosity
estimated from the $L^{\prime}$-band (3.8 $\mu$m), [O IV] 25.89 $\mu$m,
[Ne V] 14.32 $\mu$m luminosities all yield large bolometric to X-ray
luminosity ratios, $\kappa_{\rm 2-10 keV} =$ 110--410. The Eddington ratio 
is high, $\lambda_{\rm Edd} =
0.4-0.8$. The large $\kappa_{\rm 2-10 keV}$ value and high Eddington
ratio are consistent with the general trend in AGNs found by
\citet{Vasudevan2007}. These results suggests that Mrk 463E contains a rapidly
growing SMBH, whereas the nucleus is not deeply buried, unlike in
typical local ULIRGs.

The bolometric to X-ray luminosity ratio of Mrk 463W, estimated from the
$L^{\prime}$-band luminosity, shows an unusually small value, 1--3. We
suggest that 
the Eddington ratio of this AGN is small ($<10^{-3}$),
in contrast to Mrk~463E. 
We speculate that Mrk~463W might be still in an early phase of 
AGN activity triggered by the merger.

\vspace{0.2in} 

Part of this work was financially supported by the Grant-in-Aid for
Scientific Research 17K05384 (Y.U.), 15K05030 (M.I.), 15H02070, 16K05296 (Y.T.), and for JSPS Fellows for young
researchers (A.T.). C.R. acknowledges financial support from the
China-CONICYT fellowship, FONDECYT 1141218, and Basal-CATA PFB-06/2007.
Support for this work was provided by the \textit{NuSTAR} mission, a
project led by the California Institute of Technology, managed by the
Jet Propulsion Laboratory and funded by the National Aeronautics and
Space Administration. This research also made use of data obtained with
\textit{XMM-Newton}, an ESA science mission with instruments and
contributions directly funded by ESA Member States and NASA, and with
Chandra, supported by the \textit{Chandra} X-ray Observatory Center,
which is operated by the Smithsonian Astrophysical Observatory for and
on behalf of NASA.

\bibliographystyle{apj}
\bibliography{reference}

\begin{thebibliography}{}
\expandafter\ifx\csname natexlab\endcsname\relax\def\natexlab#1{#1}\fi

\bibitem[{{Anders} \& {Grevesse}(1989)}]{Anders1989}
{Anders}, E., \& {Grevesse}, N. 1989, \gca, 53, 197

\bibitem[{{Armus} {et~al.}(2004){Armus}, {Charmandaris}, {Spoon}, {Houck},
  {Soifer}, {Brandl}, {Appleton}, {Teplitz}, {Higdon}, {Weedman}, {Devost},
  {Morris}, {Uchida}, {van Cleve}, {Barry}, {Sloan}, {Grillmair}, {Burgdorf},
  {Fajardo-Acosta}, {Ingalls}, {Higdon}, {Hao}, {Bernard-Salas}, {Herter},
  {Troeltzsch}, {Unruh}, \& {Winghart}}]{Armus2004}
{Armus}, L., {Charmandaris}, V., {Spoon}, H.~W.~W., {et~al.} 2004, \apjs, 154,
  178

\bibitem[{{Arnaud}(1996)}]{Arnaud1996}
{Arnaud}, K.~A. 1996, in Astronomical Society of the Pacific Conference Series,
  Vol. 101, Astronomical Data Analysis Software and Systems V, ed. G.~H.
  {Jacoby} \& J.~{Barnes}, 17

\bibitem[{{Barvainis}(1987)}]{Barvainis1987}
{Barvainis}, R. 1987, \apj, 320, 537

\bibitem[{{Bianchi} {et~al.}(2008){Bianchi}, {Chiaberge}, {Piconcelli},
  {Guainazzi}, \& {Matt}}]{Bianchi2008}
{Bianchi}, S., {Chiaberge}, M., {Piconcelli}, E., {Guainazzi}, M., \& {Matt},
  G. 2008, \mnras, 386, 105

\bibitem[{{Bianchi} {et~al.}(2006){Bianchi}, {Guainazzi}, \&
  {Chiaberge}}]{Bianchi2006}
{Bianchi}, S., {Guainazzi}, M., \& {Chiaberge}, M. 2006, \aap, 448, 499

\bibitem[{{Dasyra} {et~al.}(2011){Dasyra}, {Ho}, {Netzer}, {Combes},
  {Trakhtenbrot}, {Sturm}, {Armus}, \& {Elbaz}}]{Dasyra2011}
{Dasyra}, K.~M., {Ho}, L.~C., {Netzer}, H., {et~al.} 2011, \apj, 740, 94

\bibitem[{{Eguchi} {et~al.}(2011){Eguchi}, {Ueda}, {Awaki}, {Aird},
  {Terashima}, \& {Mushotzky}}]{Eguchi2011}
{Eguchi}, S., {Ueda}, Y., {Awaki}, H., {et~al.} 2011, \apj, 729, 31

\bibitem[{{Fabbiano} {et~al.}(2011){Fabbiano}, {Wang}, {Elvis}, \&
  {Risaliti}}]{Fabbiano2011}
{Fabbiano}, G., {Wang}, J., {Elvis}, M., \& {Risaliti}, G. 2011, \nat, 477, 431

\bibitem[{{Gandhi} {et~al.}(2015){Gandhi}, {H{\"o}nig}, \&
  {Kishimoto}}]{Gandhi2015}
{Gandhi}, P., {H{\"o}nig}, S.~F., \& {Kishimoto}, M. 2015, \apj, 812, 113

\bibitem[{{Garmire} {et~al.}(2003){Garmire}, {Bautz}, {Ford}, {Nousek}, \&
  {Ricker}}]{Garmire2003}
{Garmire}, G.~P., {Bautz}, M.~W., {Ford}, P.~G., {Nousek}, J.~A., \& {Ricker},
  Jr., G.~R. 2003, in \procspie, Vol. 4851, X-Ray and Gamma-Ray Telescopes and
  Instruments for Astronomy., ed. J.~E. {Truemper} \& H.~D. {Tananbaum}, 28--44

\bibitem[{{Harrison} {et~al.}(2013){Harrison}, {Craig}, {Christensen},
  {Hailey}, {Zhang}, {Boggs}, {Stern}, {Cook}, {Forster}, {Giommi},
  {Grefenstette}, {Kim}, {Kitaguchi}, {Koglin}, {Madsen}, {Mao}, {Miyasaka},
  {Mori}, {Perri}, {Pivovaroff}, {Puccetti}, {Rana}, {Westergaard}, {Willis},
  {Zoglauer}, {An}, {Bachetti}, {Barri{\`e}re}, {Bellm}, {Bhalerao},
  {Brejnholt}, {Fuerst}, {Liebe}, {Markwardt}, {Nynka}, {Vogel}, {Walton},
  {Wik}, {Alexander}, {Cominsky}, {Hornschemeier}, {Hornstrup}, {Kaspi},
  {Madejski}, {Matt}, {Molendi}, {Smith}, {Tomsick}, {Ajello}, {Ballantyne},
  {Balokovi{\'c}}, {Barret}, {Bauer}, {Blandford}, {Brandt}, {Brenneman},
  {Chiang}, {Chakrabarty}, {Chenevez}, {Comastri}, {Dufour}, {Elvis}, {Fabian},
  {Farrah}, {Fryer}, {Gotthelf}, {Grindlay}, {Helfand}, {Krivonos}, {Meier},
  {Miller}, {Natalucci}, {Ogle}, {Ofek}, {Ptak}, {Reynolds}, {Rigby},
  {Tagliaferri}, {Thorsett}, {Treister}, \& {Urry}}]{Harrison2013}
{Harrison}, F.~A., {Craig}, W.~W., {Christensen}, F.~E., {et~al.} 2013, \apj,
  770, 103

\bibitem[{{Heckman} {et~al.}(2005){Heckman}, {Ptak}, {Hornschemeier}, \&
  {Kauffmann}}]{Heckman2005}
{Heckman}, T.~M., {Ptak}, A., {Hornschemeier}, A., \& {Kauffmann}, G. 2005,
  \apj, 634, 161

\bibitem[{{Ho}(2008)}]{Ho2008}
{Ho}, L.~C. 2008, \araa, 46, 475

\bibitem[{{Hopkins} {et~al.}(2008){Hopkins}, {Hernquist}, {Cox}, \& {Kere{\v
  s}}}]{Hopkins2008}
{Hopkins}, P.~F., {Hernquist}, L., {Cox}, T.~J., \& {Kere{\v s}}, D. 2008,
  \apjs, 175, 356

\bibitem[{{Ikeda} {et~al.}(2009){Ikeda}, {Awaki}, \& {Terashima}}]{Ikeda2009}
{Ikeda}, S., {Awaki}, H., \& {Terashima}, Y. 2009, \apj, 692, 608

\bibitem[{{Imanishi} \& {Saito}(2014)}]{Imanishi2014}
{Imanishi}, M., \& {Saito}, Y. 2014, \apj, 780, 106

\bibitem[{{Imanishi} \& {Terashima}(2004)}]{Imanishi2004}
{Imanishi}, M., \& {Terashima}, Y. 2004, \aj, 127, 758

\bibitem[{{Imanishi} {et~al.}(2003){Imanishi}, {Terashima}, {Anabuki}, \&
  {Nakagawa}}]{Imanishi2003}
{Imanishi}, M., {Terashima}, Y., {Anabuki}, N., \& {Nakagawa}, T. 2003, \apjl,
  596, L167

\bibitem[{{Iwasawa} {et~al.}(2011{\natexlab{a}}){Iwasawa}, {Sanders}, {Teng},
  {U}, {Armus}, {Evans}, {Howell}, {Komossa}, {Mazzarella}, {Petric}, {Surace},
  {Vavilkin}, {Veilleux}, \& {Trentham}}]{Iwasawa2011}
{Iwasawa}, K., {Sanders}, D.~B., {Teng}, S.~H., {et~al.} 2011{\natexlab{a}},
  \aap, 529, A106

\bibitem[{{Iwasawa} {et~al.}(2011{\natexlab{b}}){Iwasawa}, {Mazzarella},
  {Surace}, {Sanders}, {Armus}, {Evans}, {Howell}, {Komossa}, {Petric}, {Teng},
  {U}, \& {Veilleux}}]{Iwasawa2011-Mrk273}
{Iwasawa}, K., {Mazzarella}, J.~M., {Surace}, J.~A., {et~al.}
  2011{\natexlab{b}}, \aap, 528, A137

\bibitem[{{Jansen} {et~al.}(2001){Jansen}, {Lumb}, {Altieri}, {Clavel}, {Ehle},
  {Erd}, {Gabriel}, {Guainazzi}, {Gondoin}, {Much}, {Munoz}, {Santos},
  {Schartel}, {Texier}, \& {Vacanti}}]{Jansen2001}
{Jansen}, F., {Lumb}, D., {Altieri}, B., {et~al.} 2001, \aap, 365, L1

\bibitem[{{Kalberla} {et~al.}(2005){Kalberla}, {Burton}, {Hartmann}, {Arnal},
  {Bajaja}, {Morras}, \& {P{\"o}ppel}}]{Kalberla2005}
{Kalberla}, P.~M.~W., {Burton}, W.~B., {Hartmann}, D., {et~al.} 2005, \aap,
  440, 775

\bibitem[{{Kawamuro} {et~al.}(2016){Kawamuro}, {Ueda}, {Tazaki}, {Ricci}, \&
  {Terashima}}]{Kawamuro2016}
{Kawamuro}, T., {Ueda}, Y., {Tazaki}, F., {Ricci}, C., \& {Terashima}, Y. 2016,
  \apjs, 225, 14

\bibitem[{{Komossa} {et~al.}(2003){Komossa}, {Burwitz}, {Hasinger}, {Predehl},
  {Kaastra}, \& {Ikebe}}]{Komossa2003}
{Komossa}, S., {Burwitz}, V., {Hasinger}, G., {et~al.} 2003, \apjl, 582, L15

\bibitem[{{Kormendy} \& {Ho}(2013)}]{Kormendy2013}
{Kormendy}, J., \& {Ho}, L.~C. 2013, \araa, 51, 511

\bibitem[{{Koss} {et~al.}(2012){Koss}, {Mushotzky}, {Treister}, {Veilleux},
  {Vasudevan}, \& {Trippe}}]{Koss2012}
{Koss}, M., {Mushotzky}, R., {Treister}, E., {et~al.} 2012, \apjl, 746, L22

\bibitem[{{Koss} {et~al.}(2011){Koss}, {Mushotzky}, {Treister}, {Veilleux},
  {Vasudevan}, {Miller}, {Sanders}, {Schawinski}, \& {Trippe}}]{Koss2011}
---. 2011, \apjl, 735, L42

\bibitem[{{Liu} {et~al.}(2014){Liu}, {Wang}, {Yang}, {Zhu}, \&
  {Zhou}}]{Liu2014}
{Liu}, T., {Wang}, J.-X., {Yang}, H., {Zhu}, F.-F., \& {Zhou}, Y.-Y. 2014,
  \apj, 783, 106

\bibitem[{{Mazzarella} {et~al.}(1991){Mazzarella}, {Soifer}, {Graham},
  {Neugebauer}, {Matthews}, \& {Gaume}}]{Mazzarella1991}
{Mazzarella}, J.~M., {Soifer}, B.~T., {Graham}, J.~R., {et~al.} 1991, \aj, 102,
  1241

\bibitem[{{Mihos} \& {Hernquist}(1996)}]{Mihos1996}
{Mihos}, J.~C., \& {Hernquist}, L. 1996, \apj, 464, 641

\bibitem[{{Minezaki} \& {Matsushita}(2015)}]{Minezaki2015}
{Minezaki}, T., \& {Matsushita}, K. 2015, \apj, 802, 98

\bibitem[{{Nandra} {et~al.}(2007){Nandra}, {O'Neill}, {George}, \&
  {Reeves}}]{Nandra2007}
{Nandra}, K., {O'Neill}, P.~M., {George}, I.~M., \& {Reeves}, J.~N. 2007,
  \mnras, 382, 194

\bibitem[{{Nardini}(2017)}]{Nardini2017}
{Nardini}, E. 2017, \mnras, 471, 3483

\bibitem[{{Oda} {et~al.}(2017){Oda}, {Tanimoto}, {Ueda}, {Imanishi},
  {Terashima}, \& {Ricci}}]{Oda2017}
{Oda}, S., {Tanimoto}, A., {Ueda}, Y., {et~al.} 2017, \apj, 835, 179

\bibitem[{{Pereira-Santaella} {et~al.}(2011){Pereira-Santaella},
  {Alonso-Herrero}, {Santos-Lleo}, {Colina}, {Jim{\'e}nez-Bail{\'o}n},
  {Longinotti}, {Rieke}, {Ward}, \& {Esquej}}]{Pereira2011}
{Pereira-Santaella}, M., {Alonso-Herrero}, A., {Santos-Lleo}, M., {et~al.}
  2011, \aap, 535, A93

\bibitem[{{Ranalli} {et~al.}(2003){Ranalli}, {Comastri}, \&
  {Setti}}]{Ranalli2003}
{Ranalli}, P., {Comastri}, A., \& {Setti}, G. 2003, \aap, 399, 39

\bibitem[{{Ricci} {et~al.}(2016){Ricci}, {Bauer}, {Treister},
  {Romero-Ca{\~n}izales}, {Arevalo}, {Iwasawa}, {Privon}, {Sanders},
  {Schawinski}, {Stern}, \& {Imanishi}}]{Ricci2016}
{Ricci}, C., {Bauer}, F.~E., {Treister}, E., {et~al.} 2016, \apj, 819, 4

\bibitem[{{Ricci} {et~al.}(2017){Ricci}, {Bauer}, {Treister}, {Schawinski},
  {Privon}, {Blecha}, {Arevalo}, {Armus}, {Harrison}, {Ho}, {Iwasawa},
  {Sanders}, \& {Stern}}]{Ricci2017}
---. 2017, \mnras, 468, 1273

\bibitem[{{Rigby} {et~al.}(2009){Rigby}, {Diamond-Stanic}, \&
  {Aniano}}]{Rigby2009}
{Rigby}, J.~R., {Diamond-Stanic}, A.~M., \& {Aniano}, G. 2009, \apj, 700, 1878

\bibitem[{{Risaliti} {et~al.}(2010){Risaliti}, {Imanishi}, \&
  {Sani}}]{Risaliti2010}
{Risaliti}, G., {Imanishi}, M., \& {Sani}, E. 2010, \mnras, 401, 197

\bibitem[{{Sanders} \& {Mirabel}(1996)}]{Sanders1996}
{Sanders}, D.~B., \& {Mirabel}, I.~F. 1996, \araa, 34, 749

\bibitem[{{Satyapal} {et~al.}(2007){Satyapal}, {Vega}, {Heckman}, {O'Halloran},
  \& {Dudik}}]{Satyapal2007}
{Satyapal}, S., {Vega}, D., {Heckman}, T., {O'Halloran}, B., \& {Dudik}, R.
  2007, \apjl, 663, L9

\bibitem[{{Smith} {et~al.}(2001){Smith}, {Brickhouse}, {Liedahl}, \&
  {Raymond}}]{Smith2001}
{Smith}, R.~K., {Brickhouse}, N.~S., {Liedahl}, D.~A., \& {Raymond}, J.~C.
  2001, \apjl, 556, L91

\bibitem[{{Struck}(1999)}]{Struck1999}
{Struck}, C. 1999, \physrep, 321, 1

\bibitem[{{Str{\"u}der} {et~al.}(2001){Str{\"u}der}, {Briel}, {Dennerl},
  {Hartmann}, {Kendziorra}, {Meidinger}, {Pfeffermann}, {Reppin}, {Aschenbach},
  {Bornemann}, {Br{\"a}uninger}, {Burkert}, {Elender}, {Freyberg}, {Haberl},
  {Hartner}, {Heuschmann}, {Hippmann}, {Kastelic}, {Kemmer}, {Kettenring},
  {Kink}, {Krause}, {M{\"u}ller}, {Oppitz}, {Pietsch}, {Popp}, {Predehl},
  {Read}, {Stephan}, {St{\"o}tter}, {Tr{\"u}mper}, {Holl}, {Kemmer}, {Soltau},
  {St{\"o}tter}, {Weber}, {Weichert}, {von Zanthier}, {Carathanassis}, {Lutz},
  {Richter}, {Solc}, {B{\"o}ttcher}, {Kuster}, {Staubert}, {Abbey}, {Holland},
  {Turner}, {Balasini}, {Bignami}, {La Palombara}, {Villa}, {Buttler},
  {Gianini}, {Lain{\'e}}, {Lumb}, \& {Dhez}}]{Struder2001}
{Str{\"u}der}, L., {Briel}, U., {Dennerl}, K., {et~al.} 2001, \aap, 365, L18

\bibitem[{{Teng} {et~al.}(2005){Teng}, {Wilson}, {Veilleux}, {Young},
  {Sanders}, \& {Nagar}}]{Teng2005}
{Teng}, S.~H., {Wilson}, A.~S., {Veilleux}, S., {et~al.} 2005, \apj, 633, 664

\bibitem[{{Teng} {et~al.}(2015){Teng}, {Rigby}, {Stern}, {Ptak}, {Alexander},
  {Bauer}, {Boggs}, {Brandt}, {Christensen}, {Comastri}, {Craig}, {Farrah},
  {Gandhi}, {Hailey}, {Harrison}, {Hickox}, {Koss}, {Luo}, {Treister}, \&
  {Zhang}}]{Teng2015}
{Teng}, S.~H., {Rigby}, J.~R., {Stern}, D., {et~al.} 2015, \apj, 814, 56

\bibitem[{{Turner} {et~al.}(2001){Turner}, {Abbey}, {Arnaud}, {Balasini},
  {Barbera}, {Belsole}, {Bennie}, {Bernard}, {Bignami}, {Boer}, {Briel},
  {Butler}, {Cara}, {Chabaud}, {Cole}, {Collura}, {Conte}, {Cros}, {Denby},
  {Dhez}, {Di Coco}, {Dowson}, {Ferrando}, {Ghizzardi}, {Gianotti}, {Goodall},
  {Gretton}, {Griffiths}, {Hainaut}, {Hochedez}, {Holland}, {Jourdain},
  {Kendziorra}, {Lagostina}, {Laine}, {La Palombara}, {Lortholary}, {Lumb},
  {Marty}, {Molendi}, {Pigot}, {Poindron}, {Pounds}, {Reeves}, {Reppin},
  {Rothenflug}, {Salvetat}, {Sauvageot}, {Schmitt}, {Sembay}, {Short},
  {Spragg}, {Stephen}, {Str{\"u}der}, {Tiengo}, {Trifoglio}, {Tr{\"u}mper},
  {Vercellone}, {Vigroux}, {Villa}, {Ward}, {Whitehead}, \&
  {Zonca}}]{Turner2001}
{Turner}, M.~J.~L., {Abbey}, A., {Arnaud}, M., {et~al.} 2001, \aap, 365, L27

\bibitem[{{Vasudevan} \& {Fabian}(2007)}]{Vasudevan2007}
{Vasudevan}, R.~V., \& {Fabian}, A.~C. 2007, \mnras, 381, 1235

\bibitem[{{Veilleux} {et~al.}(2009){Veilleux}, {Rupke}, {Kim}, {Genzel},
  {Sturm}, {Lutz}, {Contursi}, {Schweitzer}, {Tacconi}, {Netzer}, {Sternberg},
  {Mihos}, {Baker}, {Mazzarella}, {Lord}, {Sanders}, {Stockton}, {Joseph}, \&
  {Barnes}}]{Veilleux2009}
{Veilleux}, S., {Rupke}, D.~S.~N., {Kim}, D.-C., {et~al.} 2009, \apjs, 182, 628

\bibitem[{{Weisskopf} {et~al.}(2002){Weisskopf}, {Brinkman}, {Canizares},
  {Garmire}, {Murray}, \& {Van Speybroeck}}]{Weisskopf2002}
{Weisskopf}, M.~C., {Brinkman}, B., {Canizares}, C., {et~al.} 2002, \pasp, 114,
  1

\end{thebibliography}

\end{document}